\documentclass[preprint,12pt]{elsarticle}

\UseRawInputEncoding
\usepackage{amsthm,amsmath,amssymb,amsfonts}
\usepackage{mathrsfs}
\biboptions{numbers,sort&compress}
\usepackage{geometry}
\usepackage[marginal]{footmisc}
\journal{Mechanical Systems and Signal Processing}

\begin{document}

\begin{frontmatter}



\title{Three-dimensional instantaneous orbit map for rotor-bearing system based on a novel multivariate complex variational mode decomposition algorithm}


\renewcommand{\thefootnote}{\fnsymbol{footnote}}
\author[1,2]{Xiaolong Cui\fnmark[3]}
\author[1]{Jie Huang\fnmark[3]}
\fntext[3]{Xiaolong Cui and Jie Huang contributed equally to this work.}
\author[1]{Chaoshun Li \corref{corresponding-author}}
\cortext[corresponding-author]{Corresponding author.}
\ead{csli@hust.edu.cn}
\author[1]{Yujie Zhao}
\address[1]{School of Civil and Hydraulic Engineering, Huazhong University of Science and Technology, Wuhan, Hubei, China, 430074}
\address[2]{Wuhan Second Ship Design and Research Institute, Wuhan, Hubei, China, 430010}
% \maketitle
\begin{abstract}
Full spectrum and holospectrum are homogenous information fusion technology developed for the fault diagnosis of rotating machinery, which is extensively exploited in the analysis of the orbits of rotor-bearing systems. However, they are not adapted for non-stationary signals, nor can they be used for fusion analysis of vibrations of multiple bearing sections. By drawing inspiration from the multivariate variational mode decomposition (MVMD) and the complex-valued signal decomposition, we propose a method called multivariate complex variational mode decomposition (MCVMD). It can simultaneously extract the forward and backward components of multiple bearing sections and realize non-stationary complex signal decomposition of multiple bearing sections of the rotor. To achieve the visualization goal of condition monitoring, we propose the three-dimensional instantaneous orbit map (3D-IOM). It enables more features of shaft vibration of a rotor system to be displayed and offers a new way for the fusion analysis of vibration signals of multiple bearing sections of rotating machinery. Furthermore, making the most of the joint information, we also provide a high-resolution time-full spectrum (Time-FS) to display the forward and backward frequency components of multiple bearing sections. The effectiveness of the proposed method through both the simulated experiment and the real-life complex-valued signals is demonstrated in this paper.
\end{abstract}

% % Graphical abstract


% %Research highlights
% \begin{highlights}
% \end{highlights}


\end{frontmatter}


\section{Introduction}
As a pivotal part of the machinery system, rotating components are widely used in mechanical equipment, such as water turbines, aero-engines, lathe, etc \citep{lei2013review,li2018entropy,al2011vibration}. In recent years, there has been an increasing interest in fault diagnosis of rotating machinery based on signal processing technology \cite{fan2008machine,bin2012early,hu2007fault}. Generally, the vibration signals of machinery are non-stationary and contain a wealth of information reflecting the state of mechanical faults \cite{ying2021permutation,zhang2021novel}. Therefore, it's essential to explore the methods for extracting fault features in nonstationary signals and apply them to the fault diagnosis of rotating machinery. 

In recent years, there has been an increasing amount of studies on the analysis of nonstationary signals, such as empirical wavelet transform (EWT) \cite{gilles2013empirical}, empirical mode decomposition (EMD) \cite{huang1998empirical}, variational mode decomposition (VMD) \cite{dragomiretskiy2013variational}. These methods are broadly employed in the field of mechanical fault diagnosis based on vibration signals \cite{li2020novel,ye2020adaptive,shi2021multistage}. However, these methods are only used to research univariate signals. Studies have shown that the fusion of homologous signals in two channels is more conducive to extracting diagnostic information of rotating machinery \cite{yu2019novel,yu2020vibration}. Full spectrum \cite{goldman1999application,han1999directional,shravankumar2016detection} can reflect the interrelation of vibration signals picked up by two probes mounted in coplanar and mutually perpendicular directions, and provide information on the precession direction. This information provided by the full spectrum is some of the key features in the fault diagnosis of rotating machinery, so the full spectrum shows great potential in the fault diagnosis of machinery \cite{zhao2012multivariate,patel2011application,jia2018coupling}. Qu et al. \cite{qu1989holospectrum} researched the rationale of holospectrum. Besides the information mentioned in the full spectrum, the holospectrum also extracts the phase information of vibration \cite{qu1989holospectrum,liangsheng1998rotor,liu2008new}. In addition, holospectrum technology can identify the underlying dangers in the operation process more accurately than general production methods. Han et al. \cite{han2011application} proposed the full vector spectrum, which is developed based on the holospectrum and the full spectrum. This technology defines the length of the semi-major axis of the rotating precessing elliptical orbit as the main vibration vector to evaluate the maximum vibration intensity of the rotor \cite{han2011application,chen2017prediction,gong2012bearing}. Although these methods have been applied to rotating machinery resoundingly, the averaging effect of the Fourier transform makes these methods unfavorable for handling non-stationary signals. In the process of acquiring the homologous signal, the two signals from the sensors in the mutually perpendicular directions of the same bearing section can be constructed into a complex signal. The complex-valued signal is a special case of a two-channel signal. Unlike the full spectrum-like two-channel signal analysis methods described above, some studies extended the univariate nonstationary signal processing technology to bivariate to handle the complex-valued signal. Rilling et al. \cite{rilling2007bivariate} raised bivariate EMD (BEMD), which directly applies the basic principles of EMD to the bivariate framework. It is widely used in image fusion \cite{rehman2009bivariate}, mechanical fault diagnosis \cite{yang2011bivariate}, and so on. But it only works for the signals associated with the Cartesian coordinate system. Furthermore, it cannot achieve the separation of positive and negative frequency signals. Different from the main idea of BEMD, the complex EMD put forward by Tanaka \cite{tanaka2007complex} cleverly applied EMD to the positive and negative components respectively, and it has shown advantages in information fusion. However, CEMD cannot guarantee the same number of intrinsic mode functions (IMFs) in the real and imaginary data channels. Inspired by the CEMD, Wang et al. \cite{wang2017complex} raised the complex VMD (CVMD). CVMD inherits the characteristics of VMD, including stronger anti-noise ability and a solid theoretical foundation. Also, CVMD has better decomposition performance than CEMD. However, these processing techniques mentioned above are only appropriate for processing signals of a single bearing section.

With the development of large-scale mechanical equipment, a single sensor is no longer sufficient to monitor the overall condition of the rotor-bearing system. Multiple sensors are needed to obtain operating information of each component of a rotor system. The advancement of multiple sensor-based condition monitoring has contributed to the flourishing of multivariate signal processing methods. For example, aside from BEMD and CEMD, methods based on EMD also include trivariate EMD (TEMD) \cite{ur2009empirical} and multivariate EMD (MEMD). TEMD method is based on extreme value calculation. As an extension of the BEMD and TEMD concept, MEMD \cite{rehman2010multivariate} processes multivariate input signals directly in the $n$-dimensional domain where the signals reside. However, the EMD-based extension methods have a prominent disadvantage, that is, the mode mixing problem. In addition, MEMD's effect depends on the appropriate number and direction of projections. Based on the same extension principle, Ahrabian and Rehaman presented the synchrosqueezing-based time-frequency analysis of multivariate data (MSST) \cite{ahrabian2015synchrosqueezing} and MVMD \cite{ur2019multivariate}, respectively. Both models supposed a common oscillation that was best suited for all the single-channel oscillations, in other words, there was a joint frequency component in multiple channels. MSST is a wavelet-based method. And from the perspective of the error bound, it displayed the possibility of establishing a localized multivariate time-frequency representation. However, MSST performed poorly in the circumstance of processing intense noise signals. MVMD is an extension of VMD, therefore, it maintains the properties of VMD, including robustness and anti-mode-mixing. Notably, it shows the properties of mode alignment. However, these multivariate signal processing techniques are only developed to handle real-valued signals, which are subjected to limitations in the field of complex-valued signal processing.

The dynamic responses in each orientation of multiple bearing sections of the rotor may differ during operation and these responses have important implications for fault diagnosis. Therefore, for rotating machinery with multiple bearing sections, it is necessary to analyze the influence of the vibration response of multiple bearing sections on the unit as a whole. Combining the properties of the complex field with the basic principles of MVMD, we propose the multivariate complex variational mode decomposition (MCVMD) algorithm. In fact, MCVMD extends MVMD to the complex domain, which is used to handle signals picked up from mutually perpendicular proximity probes mounted on multiple bearing sections. The proposed method adequately integrates the information of multiple bearing sections. In addition, accurate acquisition of the instantaneous characteristics of the signal is the key to condition monitoring and fault diagnosis. In our past work \cite{cui2020instantaneous}, we investigated instantaneous feature extraction methods for precession elliptical orbits. As an extension of this work, this paper proposes an instantaneous feature extraction method for precession elliptical orbits of multiple bearing sections. Based on the instantaneous features of multiple bearing sections, we propose 3D-IOM and Time-FS. The 3D-IOM is utilized for exhibiting the instantaneous vibration state of the rotor-bearing system. Since the frequencies of vibration components are consistent, the vibration response of the rotor-bearing system is interrelated in each bearing section. Thus, the 3D-IOM makes full use of the correlation of the complex-valued signals across multiple bearing sections. It can give the amplitude, phase, and frequency information of unit vibration comprehensively.

The overall structure of the study is as follows. Section \ref{Section2} draws a description of the theoretical background, including full spectrum and MVMD algorithms. Section \ref{section3} describes the proposed method MCVMD in detail. In Section \ref{section4}, several experiments are made employing the proposed method to verify the effectiveness. The conclusions are given in Section \ref{section5}.

\section{Theoretical background}\label{Section2}
\subsection{Full spectrum}
In full spectrum, the signals picked up by two probes mounted in coplanar and mutually perpendicular directions are needed, one in the horizontal direction $(x)$ and the other in the vertical direction $(y)$. There is a waveform in each sensor. After combining these two waveforms, a direct orbit is formed, which is the sum of the filtered orbits. Because the relative phase information between the $X$ spectrum and the $Y$ spectrum is lost, the half spectrum cannot display the rotation direction and precession information of the rotor \cite{goldman1999application}. The full spectrum can present the precession direction and the orbit ellipticity simultaneously. Actually, each filtered orbit is elliptical because it is a superposition of two circular orbits\cite{bachschmid2004diagnostic,lee1998directional}. The two rotation vectors that make up the circular orbit rotate in the opposite directions with radius of ${R_{\omega {\rm{ + }}}}$ (forward) and ${R_{\omega {\rm{ - }}}}$ (backward) at the same frequency $ \omega $. The forward response vector is ${R_{\omega {\rm{ + }}}}{e^{j(\omega t + {\alpha _\omega })}}$, while the backward response vector is ${R_{\omega  - }}{e^{ - j(\omega t + {\beta _\omega })}}$. The instantaneous position of the rotor on the filtered orbit is the sum of these two vectors, where ${\alpha _\omega }$ and ${\beta _\omega }$ are phases of forward and backward responses, respectively. Note that the major axis of the filtered elliptical orbit is ${R_{\omega  + }} + {R_{\omega  - }}$, while its minor axis is $\left| {{R_{\omega  + }} - {R_{\omega  - }}} \right|$. Forward precession means that the rotor precession direction is the same as the rotation direction, i.e. ${R_{\omega  + }} > {R_{\omega  - }}$. Conversely, backward precession means that the rotor precession direction is opposite to the rotation direction, i.e. ${R_{\omega  + }} < {R_{\omega  - }}$. In addition, the inclination angle between the major axis and the horizontal probe is needed, which can be utilized to describe an ellipse thoroughly. The inclination angle is defined as $\left( {{\beta _\omega } + {\alpha _\omega }} \right)/2$.

The forward and backward amplitudes are given as:
\begin{equation}
    \left\{ {\begin{array}{*{20}{c}}
        {{R_{\omega  + }} = \sqrt {X_\omega ^2 + Y_\omega ^2 + 2{X_\omega }{Y_\omega }\sin \left( {{\phi _{x\omega }} - {\phi _{y\omega }}} \right)} }\\
        {{R_{\omega  - }} = \sqrt {X_\omega ^2 + Y_\omega ^2 - 2{X_\omega }{Y_\omega }\sin \left( {{\phi _{x\omega }} - {\phi _{y\omega }}} \right)} }
        \end{array}} \right.,
\end{equation}
where ${X_\omega }$ and ${Y_\omega }$ are the magnitude of the signal $x$ and $y$, respectively. ${\phi _{x\omega }}$ and ${\phi _{y\omega }}$ are the initial phases of the signal $x$ and $y$, respectively.
\subsection{Multivariate Variational Mode Decomposition}
MVMD first represents a set of $C$ real-valued amplitude modulated-frequency modulated (AM-FM) signals in a vector form
\begin{equation}
    {\bf{u}}(t) = \left[ {\begin{array}{*{20}{c}}
        {{u_1}(t)}\\
        {{u_2}(t)}\\
         \vdots \\
        {{u_C}(t)}
        \end{array}} \right] = \left[ {\begin{array}{*{20}{c}}
        {{a_1}\cos ({\phi _1}(t))}\\
        {{a_2}\cos ({\phi _2}(t))}\\
         \vdots \\
        {{a_C}\cos ({\phi _C}(t))}
        \end{array}} \right],
\end{equation}
where ${a_i}(t)$ and ${\phi _i}(t)$ represent amplitude and phase function corresponding to the $i$-th signal component, respectively. Next, MVMD obtains an analytic representation of the AM-FM signal by employing the Hilbert transform operator as:
\begin{equation}
    \begin{array}{c}
        {{\bf{u}}_ + }(t) = {\bf{u}}(t) + j{\rm{{\cal H}}}({\bf{u}}(t))\\
         = \left[ {\begin{array}{*{20}{c}}
        {u_{_ + }^1(t)}\\
        {u_ + ^2(t)}\\
         \vdots \\
        {u_ + ^C(t)}
        \end{array}} \right] = \left[ {\begin{array}{*{20}{c}}
        {{a_1}{e^{j{\phi _1}(t)}}}\\
        {{a_2}{e^{j{\phi _2}(t)}}}\\
         \vdots \\
        {{a_C}{e^{j{\phi _C}(t)}}}
        \end{array}} \right].
        \end{array}
    \end{equation}
where ${\rm{{\cal H}}}( \cdot )$ denotes the operator of Hilbert Transform. 

The real-valued signal is utilized in MVMD, and the form of it is obtained as ${\bf{u}}(t) = \Re \{ {{\bf{u}}_ + }(t)\}$, where $\Re ( \cdot )$ is the operator of extracting the real part. MVMD assumes a single common component across multiple channels, therefore, the resulting formula is given by
\begin{equation}
    {{\bf{u}}_ + }(t) = \left| {{{\bf{u}}_ + }(t)} \right|{e^{j\phi (t)}}.
    \end{equation}

The goal of MVMD is to decompose the input data ${\bf{x}}(t)$ of $C$ channels into predefined $k$ IMFs ${{\bf{u}}_k}(t)$, i.e., ${\bf{x}}(t) = \left[ {{x_1}(t),{x_2}(t), \cdots {x_C}(t)} \right]$ 
\begin{equation}
    {\bf{x}}(t) = \sum\limits_{k = 1}^K {{{\bf{u}}_k}(t)},
\end{equation}
where ${{\bf{u}}_{\bf{k}}}(t) = [{u_1}(t),{u_2}(t), \cdots {u_C}(t)].$

The set of IMFs $\left\{ {{{\bf{u}}_k}(t)} \right\}_{k = 1}^K$ in input data are selected by a variational model that intends to minimize the sum of bandwidth and reconstruct the input signal accurately at the same time. To obtain the minimum bandwidth, MVMD starts from getting the analytical signal ${\bf{u}}_ + ^k(t)$. Next, shifting the ${\bf{u}}_ + ^k(t)$ harmonically by $\omega $. The bandwidth is now estimated by using the ${L_2}$ norm of the gradient function of the harmonically shifted ${\bf{u}}_ + ^k(t)$. 

Notably, the entire vector ${\bf{u}}_ + ^k(t)$ uses a single common frequency component in the harmonic mixing. Thus, MVMD expects to find an ensemble of IMFs in ${{\bf{u}}_k}(t)$ with a single common frequency component ${\omega _k}$ across multiple channels. The resulting cost function $f$ for MVMD is given by 
\begin{equation}
    f = \sum\limits_k {\sum\limits_c {\left\| {{\partial _t}\left[ {u_ + ^{k,c}(t){e^{ - j{\omega _k}t}}} \right]} \right\|_2^2} }, 
\end{equation}
where $u_ + ^{k,c}$ denotes the analytic modulated signal corresponding to channel $c$ and mode $k$. The related constrained optimization problem for MVMD becomes:
\begin{align}
    \begin{array}{*{20}{c}}
        {\mathop {{\rm{min}}}\limits_{\left\{ {{u_{k,c}}} \right\},\left\{ {{\omega _k}} \right\}} \left\{ {\sum\limits_k {\sum\limits_c {\left\| {{\partial _t}\left[ {u_ + ^{k,c}(t){e^{ - j{\omega _k}t}}} \right]} \right\|_2^2} } } \right\}}\\
        {{\kern 1pt} {\kern 1pt} {\kern 1pt} {\kern 1pt} {\kern 1pt} {\kern 1pt} {\kern 1pt} {\kern 1pt} {\kern 1pt} {\kern 1pt} {\kern 1pt} {\kern 1pt} {\kern 1pt} {\kern 1pt} {\kern 1pt} {\kern 1pt} {\rm{s}}{\rm{.t}}{\rm{.  }}\sum\limits_k {{u_{k,c}}(t) = {x_c}(t),c = 1,2, \cdots ,C.} }
        \end{array}.
\end{align}

Next, the corresponding augmented Lagrangian function is given below
\begin{equation}
    \begin{array}{l}
        {\rm{{\cal L}}}\left( {\left\{ {{u_{k,c}}} \right\}{\rm{,}}\left\{ {{\omega _k}} \right\},{\lambda _c}} \right) = \alpha \sum\limits_k {\sum\limits_c {\left\| {{\partial _t}\left[ {u_ + ^{k,c}(t){e^{ - j{\omega _k}t}}} \right]} \right\|_2^2} } \\
         + \sum\limits_c {\left\| {{x_c}(t) - \sum\limits_k {{u_{k,c}}(t)} } \right\|_2^2}  + \sum\limits_c {\left\langle {{\lambda _c}(t),{x_c}(t) - \sum\limits_k {{u_{k,c}}(t)} } \right\rangle .} 
        \end{array}
\end{equation}

The unconstrained optimization problem above is solved using the alternating direction method of multipliers approach. Following, the attention is paid to the update of the modes $\hat u_{k,c}^{n + 1}(\omega )$ and the center frequencies $\omega _k^{n + 1}$ 
 \begin{equation}
    \hat u_{k,c}^{n + 1}(\omega ) = \frac{{{{\hat x}_c}(\omega ) - \sum\nolimits_{i \ne k} {{{\hat u}_{i,c}}(\omega ) + \frac{{{{\hat \lambda }_c}(\omega )}}{2}} }}{{1 + 2\alpha {{(\omega  - {\omega _k})}^2}}},
 \end{equation}
 \begin{equation}
    \omega _k^{n + 1} = \frac{{\sum\nolimits_c {\int_0^\infty  {\omega {{\left| {{{\hat u}_{k,c}}(\omega )} \right|}^2}d\omega } } }}{{\sum\nolimits_c {\int_0^\infty  {{{\left| {{{\hat u}_{k,c}}(\omega )} \right|}^2}d\omega } } }}.
 \end{equation}
While updating ${\omega _k}$ for each mode in MVMD, contributions from the power spectrum of all the $C$ channels are considered.

The Lagrange multiplier is updated by
\begin{equation}
    \hat \lambda _c^{n + 1}(\omega ) = \hat \lambda _c^n(\omega ) + \tau \left( {{{\hat x}_c}(\omega ) - \sum\limits_k {\hat u_{k,c}^{n + 1}(\omega )} } \right).
\end{equation}
\section{Proposed method}\label{section3}
\subsection{Multivariate complex variational mode decomposition}
MVMD is developed for only real signals, it doesn't work while confronted with complex-valued signals. Thus, the MCVMD algorithm is introduced in this section to account for the multivariate complex-valued non-stationary signals. Now, we consider a set of signals on multiple bearing sections of a rotor-bearing system, where the signals at each bearing section are measured by sensors placed in mutually perpendicular directions. Now, we give a set of complex-valued signals of $K$ channel, i.e. ${\rm{P}}(t) = [{p_1}(t),{p_2}(t), \cdots {p_K}(t)]$ 
\begin{equation}
    {p_i}(t) = {x_i}(t) + j{y_i}(t){\rm{    }},i{\rm{ = 1,2,}} \cdots {\rm{,}}K,
    \label {eq12}
\end{equation}
where ${x_i}(t)$ and ${y_i}(t)$ are both real signals, $j$ denotes the imaginary number. 

Generally, the complex-valued signal ${p_i}(t)$ is associated with a moving point, or a moving vector drawn from the origin, in the plane whose Cartesian coordinates are ${x_i}(t)$ and ${y_i}(t)$. The locus of the moving vectors can be expressed as two circles moving in the opposite direction with the same speed $\omega $. Based on that fact, the term ${p_i}(t)$ can be expressed in another form 
\begin{equation}
    {p_i}(t) = {p_i}^f(t) + {p_i}^b(t) = {r_i}^f{e^{j\omega t}} + {r_i}^b{e^{ - j\omega t}}{\rm{  (}}i = 1,2, \cdots ,K{\rm{)}},
\end{equation}
where ${r_i}^f = |{r_i}^f|{e^{j{\phi _i}^f}}$, ${r_i}^b = |{r_i}^b|{e^{j{\phi _i}^b}}$. Here, the superscripts $f$ and $b$ denote the forward (counter-clockwise) and backward (clockwise) frequency component. 

In signal processing, the analytic signal contains the original function and its Hilbert transform. In \cite{lee1998directional}, the forward and backward components linked with $p(t)$ is defined as 
\begin{equation}
    {p^f}(t) = {{\left[ {p(t) + j{\rm{{\cal H}}}\left( {p(t)} \right)} \right]} \mathord{\left/
 {\vphantom {{\left[ {p(t) + j{\rm{{\cal H}}}\left( {p(t)} \right)} \right]} 2}} \right.
 \kern-\nulldelimiterspace} 2},{p^b}(t) = {{\left[ {p(t) - j{\rm{{\cal H}}}\left( {p(t)} \right)} \right]} \mathord{\left/
 {\vphantom {{\left[ {p(t) - j{\rm{{\cal H}}}\left( {p(t)} \right)} \right]} 2}} \right.
 \kern-\nulldelimiterspace} 2}.
\end{equation}
The corresponding Fourier transforms, ${P^f}(\omega )$ and ${P^b}(\omega )$, are denoted as:
\begin{equation}
    {P^f}(\omega ) = [P(\omega ) + {\mathop{\rm sgn}} (\omega )P(\omega )]/2 = \left\{ {\begin{array}{*{20}{c}}
        {P(\omega ),{\rm{      for }}\omega {\rm{ > 0,}}}\\
        {P(\omega )/2,{\rm{ for }}\omega {\rm{ = 0,}}}\\
        {0,{\rm{            for  \omega  < 0}}}
        \end{array}} \right\},
\end{equation}
\begin{equation}
    {P^b}(\omega ) = [P(\omega ) - {\mathop{\rm sgn}} (\omega )P(\omega )]/2 = \left\{ {\begin{array}{*{20}{c}}
        {0,{\rm{             for }}\omega {\rm{ > 0,}}}\\
        {P(\omega )/2,{\rm{ for }}\omega {\rm{ = 0,}}}\\
        {P(\omega ),{\rm{      for  \omega  < 0}}{\rm{.}}}
        \end{array}} \right\}.
\end{equation}
where ${\mathop{\rm sgn}} $ is short for sign function, which is defined as:
\begin{equation}
    {\mathop{\rm sgn}} (\omega ) = \left\{ \begin{array}{c}
        1,\omega  > 0\\
        0,\omega  = 0\\
         - 1,\omega  < 0
        \end{array} \right..
\end{equation}

According to the definition of the analytic signal, the ${P^f}(\omega )$ and the complex conjugate of the ${P^b}(\omega )$ are both analytic signals. Due to the properties of the analytical signals, one can only handle real parts of ${p^f}(t)$ and ${p^b}(t)$ without the loss of information. Therefore, we extract the real part of ${p^f}(t)$ and ${p^b}(t)$, i.e., ${p_i}_ + (t)$ and ${p_{i - }}(t)$, respectively 
\begin{equation}
    {p_{i + }}(t) = \Re \{ {p_i}^f(t)\} ,{p_{i - }}(t) = \Re \{ {p_i}^b(t)\} ,
    \label{eq20}
\end{equation}
where $p_i^f(t)$ and $p_i^b(t)$ denotes the forward and backward component of $i$-th channel. Here, MVMD is used to decompose the ${p}_ + (t)$ and ${p}_ - (t)$ of multiple channels.

On account of the properties of the MVMD, MCVMD guarantees the same number of IMFs across multiple channels. Assuming that there are $N$ modes in each bearing section, the complex-valued signal in each channel (i.e., each bearing section) can be decomposed into forward and backward components. After the treatment of MVMD, the forward and backward components of channel $i$ are formulated as
\begin{equation}
    {p_{i + }}(t) = \sum\limits_{n = 1}^N {x_{n,i}^ + } ,{p_{i - }}(t) = \sum\limits_{n = 1}^N {x_{_{n,i}}^ - } ,
\end{equation}

Further, the reconstructed signal ${p_i}(t)$ is expressed as:
\begin{equation}
    {p_i}(t) = ({p_{i + }}(t) + j{\rm{{\cal H}}}({p_{i + }}(t))) + {({p_{i - }}(t) + j{\rm{{\cal H}}}({p_{i - }}(t)))^ * },
\end{equation}
where $*$ denotes complex conjugate operation. Also, we give another form of ${p_i}(t)$ as:
\begin{equation}
    {p_i}(t) = p_i^f(t) + p_i^b(t) = \sum\limits_{n = 1}^N {(z_{n,i}^f(t) + z_{n,i}^b(t))} ,
    \label{eq23}
\end{equation}
\begin{equation}
    z_{n,i}^f = x_{n,i}^ + (t) + j{\rm{{\cal H}}}(x_{n,i}^ + (t)),z_{n,i}^b = (x_{n,i}^ - (t) + j{\rm{{\cal H}}}{(x_{n,i}^ - (t))^ * },
    \label{eq24}
\end{equation}
where, $z_{n,i}^f(t)$ and $z_{n,i}^b(t)$ are defined as the $n$-th complex-valued IMF of forward component and backward component of $i$-th channel respectively.
\subsection{3D-IOM and time-FS}
To intuitively reveal and analyze the instantaneous vibration state of multiple bearing sections of a rotor-bearing system, Qu et al. \cite{qu1989holospectrum} developed the three-dimensional holospectrum. The three-dimensional holospectrum presents the ellipse orbit of each order of different bearing sections which can be used to analyze the state of the force bearing of the rotor. Thus, it provides a new idea for vibration analysis of multiple bearing sections of a rotor system. But the three-dimensional holospectrum is based on the Fourier transform, which shows the average state of the rotor over a period. As we all know, the transient state can better reveal the mechanical fault. Therefore, we constructed the 3D-IOM to present the instantaneous vibration state of the rotor-bearing system.

The instantaneous amplitudes of the forward and backward components can be provided by their complex-valued IMFs, which can be expressed as:
\begin{equation}
    \left\{ {\begin{array}{*{20}{c}}
        {R_{n,i}^ + (t) = \left| {z_{n,i}^f(t)} \right|}\\
        {R_{n,i}^ - (t) = \left| {z_{n,i}^b(t)} \right|}
        \end{array}} \right..
\end{equation}

As was mentioned in \cite{goldman1999application}, the instantaneous axis orbit of the components is an ellipse. According to the geometric analysis, the modulus of the resulting vector is maximum when the forward and the backward component vectors coincide. In this case, the two vectors are added to obtain the semi-major axis. The modulus of the resulting vector is smallest when the directions of the forward and backward component vectors differ by 180°. At this point, the two vectors are subtracted to obtain the semi-minor axis. The semi-major axis and the semi-minor axis of the ellipse can be obtained by
\begin{equation}
    \left\{ {\begin{array}{*{20}{c}}
        {R_a^{n,i}(t) = R_{n,i}^ + (t) + R_{n,i}^ - (t)}\\
        {R_b^{n,i}(t) = \left| {R_{n,i}^ + (t) - R_{n,i}^ - (t)} \right|}
        \end{array}} \right..
\end{equation}

Next, we rewrite the complex IMFs of the forward and backward components into complex exponential form as
\begin{equation}
    \left\{ \begin{array}{l}
        z_{n,i}^f(t) = R_{n,i}^ + (t)*{e^{j\phi _{n,i}^f\left( t \right)}}\\
        z_{n,i}^b(t) = R_{n,i}^ - (t)*{e^{j\phi _{n,i}^b\left( t \right)}}
        \end{array} \right.,
\end{equation}
where $\phi _{n,i}^f$ and $\phi _{n,i}^b$ denote the instantaneous phases of forward and backward components, respectively. According to the property of complex exponential function, the value of the instantaneous phase can be easily acquired by arc-tangent function as
\begin{equation}
    \phi _{n,i}^f\left( t \right) = \arctan \left( {\frac{{\Im (z_{n,i}^f(t))}}{{\Re (z_{n,i}^f(t))}}} \right),\phi _{n,i}^b\left( t \right) = \arctan \left( {\frac{{\Im (z_{n,i}^b(t))}}{{\Re (z_{n,i}^b(t))}}} \right),
\end{equation}
where $\Im ( \cdot )$ denotes the operator of extracting the imaginary part. Further, according to the full spectrum, the instantaneous phases of forward and backward components can also be expressed in the form that
\begin{equation}
    \phi _{n,i}^f\left( t \right) = \int_0^t {{\omega _{n,i}}(\tau )d\tau }  + {\alpha _{n,i}}\left( t \right),\phi _{n,i}^b\left( t \right) = \int_0^t { - {\omega _{n,i}}(\tau )d\tau }  + {\beta _{n,i}}(t),
\end{equation}
where, ${\omega _{n,i}}(\tau )$ represents the instantaneous angular velocity of the component vector. 

The angle between the major axis and the horizontal axis is the inclination of the major axis. Now, the instantaneous inclination angle of the orbit can be given by 
\begin{equation}
    {\theta _{n,i}}(t) = \frac{{{\alpha _{n,i}}\left( t \right) + {\beta _{n,i}}(t)}}{2} = \frac{{\phi _{n,i}^f\left( t \right) + \phi _{n,i}^b\left( t \right)}}{2}.
\end{equation}

In \cite{lee1998directional}, a shape and directivity index (SDI) is developed to measure the shape and direction information of an orbit. Now, we refer to the form in paper \cite{cui2020instantaneous}, and propose a new computing method of SDI:
\begin{align}
    {\rm{SD}}{{\rm{I}}_{n,i}}(t) =& \sin \left( {{\phi _{n,ix}}\left( t \right) - {\phi _{n,iy}}\left( t \right)} \right) \notag \\
    =& \frac{{R{{_{n,i}^ + }^2} - R{{_{n,i}^ - }^2}}}{{\left| {\Re ({p_{n,i}}(t)) + j{\cal H}\left( {\Re ({p_{n,i}}(t))} \right)} \right| \times \left| {\Im ({p_{n,i}}(t)) + j{\cal H}\left( {\Im ({p_{n,i}}(t))} \right)} \right|}},
\end{align}
where ${p_{n,i}}(t) = z_{n,i}^f(t) + z_{n,i}^b(t)$. 

\begin{figure}
    \centering
    \includegraphics[width=0.7\linewidth]{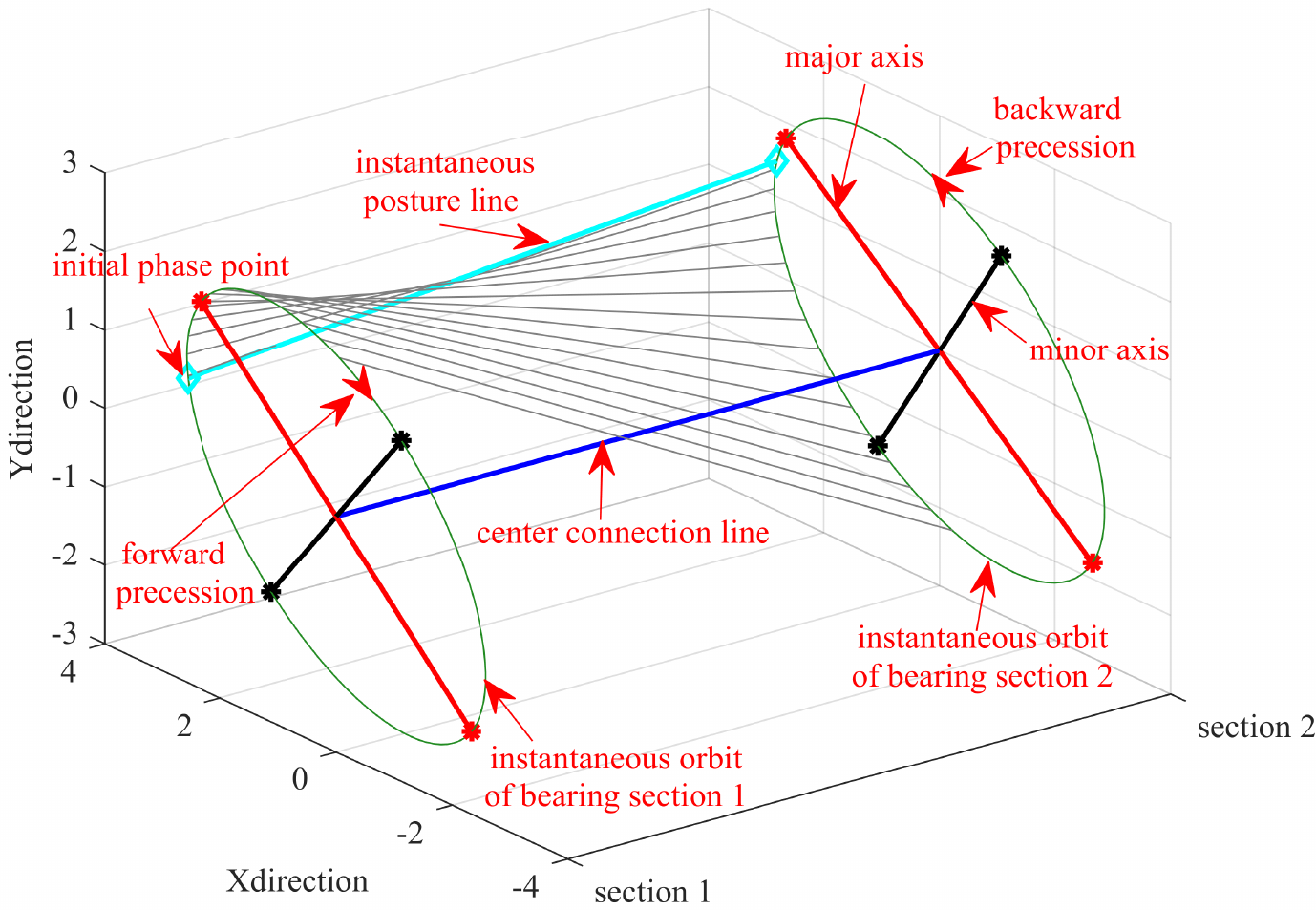}
    \caption{Sketch drawing of 3D-IOM of rotor vibration of rotating machinery.}
    \label{figure34}
\end{figure}

As derived above, the instantaneous orbit of multiple bearing sections at any time can be determined. Based on the features of the instantaneous orbit proposed above, we provide a 3D-IOM to represent the instantaneous vibration state of multiple bearing sections of the rotor. Similar to the generating line in the three-dimensional holospectrum, we use the instantaneous posture line to connect the corresponding points between different bearing sections. The drawing method is as follows. First, we find the initial phase points, and then we use the instantaneous posture line to connect the initial phase points on different bearing sections. Next, we take the initial phase point as the starting point and rotate the instantaneous posture line at equal angles according to the precession directions of the instantaneous orbit of different bearing sections. Fig. \ref{figure34} shows the diagram of the 3D-IOM.

Next, we present a time-frequency representation method (i.e. Time-FS) for multivariate vibration signals of a rotor-bearing system. The first step is projecting the instantaneous features of the components on each bearing section onto the time-frequency coordinates. Because multiple channels of the component signal share the common frequency, the corresponding instantaneous frequency (IF) can be approximated by
\begin{equation}
    {\omega _n}(t) = \frac{{\sum\limits_{i = 1}^K {\left( {{{d\phi _{n,i}^f\left( t \right)} \mathord{\left/
 {\vphantom {{d\phi _{n,i}^f\left( t \right)} {dt}}} \right.
 \kern-\nulldelimiterspace} {dt}} - {{d\phi _{n,i}^b\left( t \right)} \mathord{\left/
 {\vphantom {{d\phi _{n,i}^b\left( t \right)} {dt}}} \right.
 \kern-\nulldelimiterspace} {dt}}} \right)} }}{{2K}}.
\end{equation}

\begin{figure}
    \centering
    \includegraphics[width=1.0\linewidth]{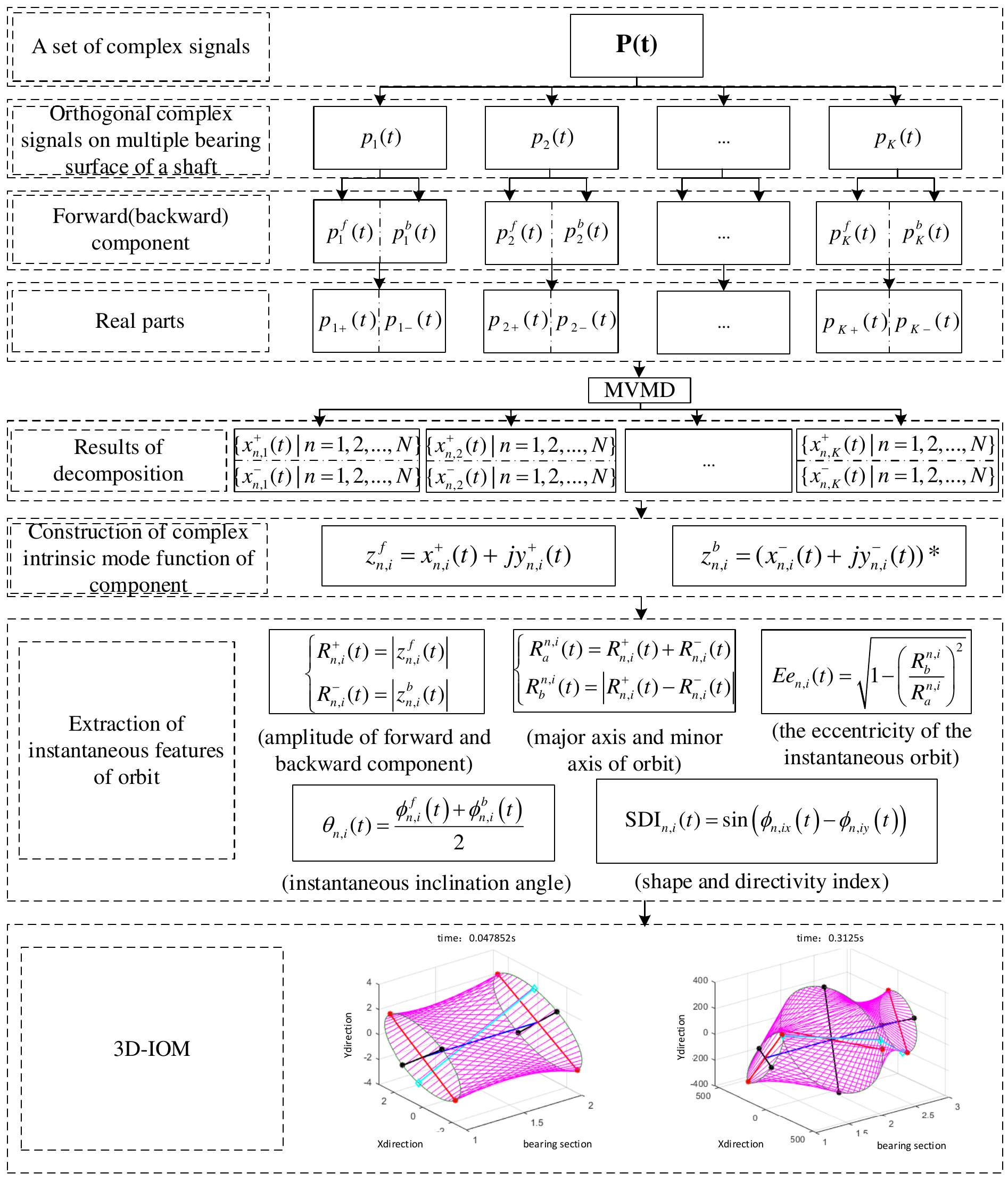}
    \caption{Flowchart of the MCVMD and 3D-IOM construction.}
    \label{figure1}
\end{figure}

Notably, we are supposed to incorporate the backward frequency components into the sum, considering the paired occurrence of the forward and backward frequency components of the rotor vibration. Correspondingly, the amplitudes of the forward component and backward component are projected onto the time-frequency plane respectively, the formula is given by:
\begin{equation}
    TF{R_{n,i}}(t,\omega ) = R_{n,i}^ + (t)\delta \left[ {\omega  - {\omega _n}(t)} \right] + R_{n,i}^ - (t)\delta \left[ { - \omega  - {\omega _n}(t)} \right].
\end{equation}
Furthermore, the time-frequency representation of the $i$-th channel on the rotor-bearing section is obtained by superposing the time-frequency representations of the $N$ modes
\begin{equation}
    TF{R_i}(t,\omega ) = \sum\limits_n {R_{n,i}^ + (t)\delta \left[ {\omega  - {\omega _n}(t)} \right] + R_{n,i}^ - (t)\delta \left[ { - \omega  - {\omega _n}(t)} \right]} .
\end{equation}
Finally, the time-frequency representations of multiple bearing sections are presented in a unified coordinate system.

Fig.\ref{figure1} illustrates the MCVMD and 3D-IOM construction process. 
\section{Experimental results}\label{section4}
To evaluate the effectiveness of the method, we will adopt the proposed method to analyze a series of test signals in this section. 
\subsection{Simulation experiment}
The first simulated signal imitated the vibration response of two bearing sections of a rotating machine during relatively smooth operation, with each bearing section containing two nonstationary components. The amplitude of the vibrations varied with time and exhibited amplitude modulation. The IFs of the two components are 16 Hz and 32 Hz, respectively. Here, the base frequency (or rotating frequency) is 16 Hz. Both sections of the signal contain 16 and 32 Hz frequency components. The sampling frequency is 1024 Hz, and the time duration of the signal is 1s. The signals are given as follows.
\begin{equation}
    \left\{ \begin{array}{l}
        {x_1}(t) = \left( {2 + 0.5 \times \cos (2.5\pi t)} \right) \times \cos \left( {2\pi  \times 16t} \right) + \left( {1.2 + 0.3 \times \cos (8\pi t)} \right) \times \cos \left( {2\pi  \times 32t} \right)\\
        {y_1}(t) = \left( {2 + 0.8 \times \cos (5\pi t)} \right) \times \cos \left( {2\pi  \times 16t + \frac{{5\pi }}{3}} \right)\\
        {\rm{       }} + \left( {1.4 + 0.53 \times \cos (6\pi t)} \right) \times \cos \left( {2\pi  \times 32t + \frac{{{\rm{2}}\pi }}{7}} \right)\\
        {x_2}(t) = \left( {2.6 + 0.7 \times \cos (5\pi t)} \right) \times \cos \left( {2\pi  \times 16t} \right) + \left( {1.5 + 0.5 \times \cos (15\pi t)} \right) \times \cos \left( {2\pi  \times 32t} \right)\\
        {y_2}(t) = \left( {2.8 + 0.6 \times \cos (10\pi t)} \right) \times \cos \left( {2\pi  \times 16t + \frac{{{\rm{3}}\pi }}{{\rm{8}}}} \right)\\
        {\rm{       }} + \left( {1.7 + 0.33 \times \cos (10\pi t)} \right) \times \cos \left( {2\pi  \times 32t + \frac{{{\rm{12}}\pi }}{7}} \right)
        \end{array} \right.
\end{equation}

\begin{figure}
    \centering
    \includegraphics[width=0.8\linewidth]{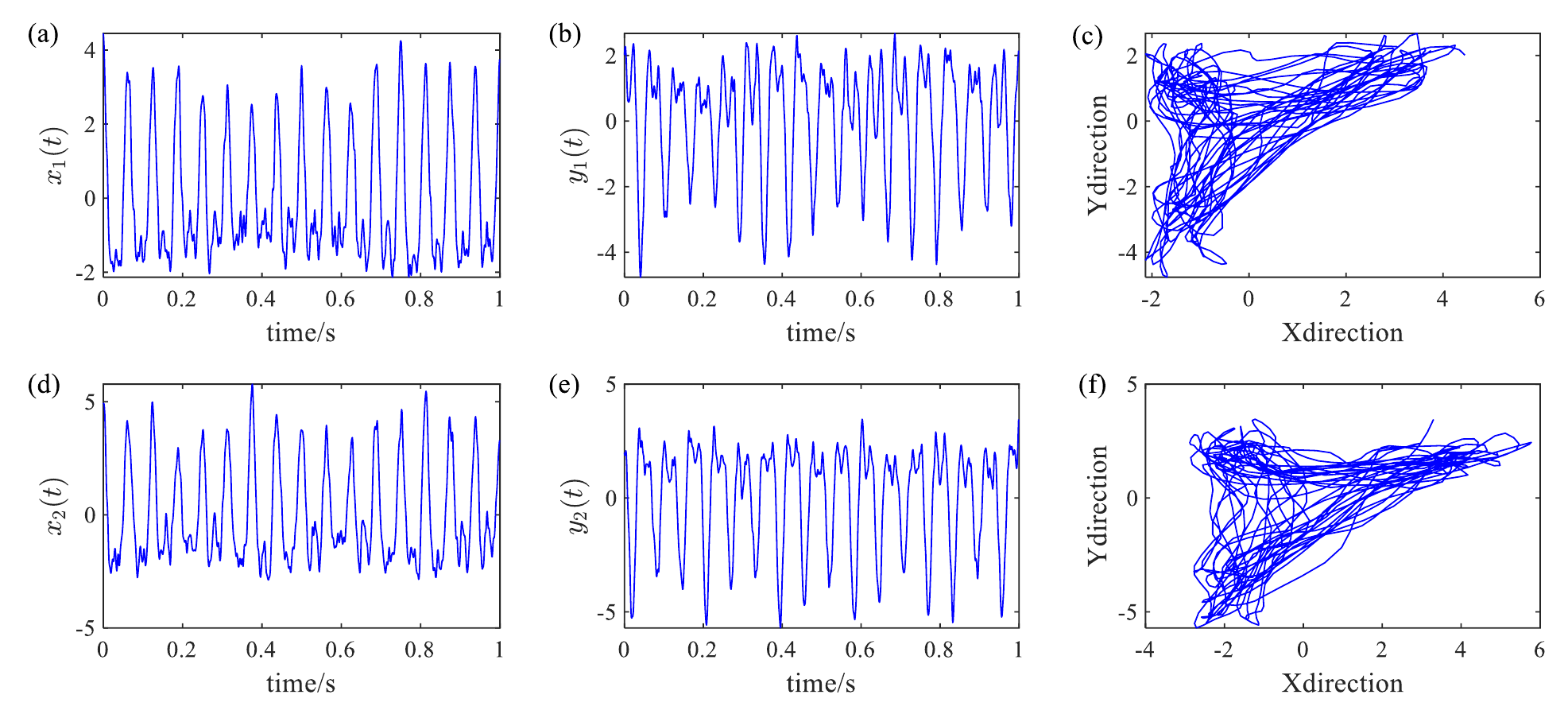}
    \caption{The noisy signal waveform and orbit. (a) and (b) Real and imaginary parts of bearing section 1. (c) Orbit of the bearing section 1. (c) and (d) Real and imaginary parts of bearing section 2. (c) Orbit of the bearing section 2.}
    \label{figure2}
\end{figure}

\begin{figure}
    \centering
    \includegraphics[width=0.8\linewidth]{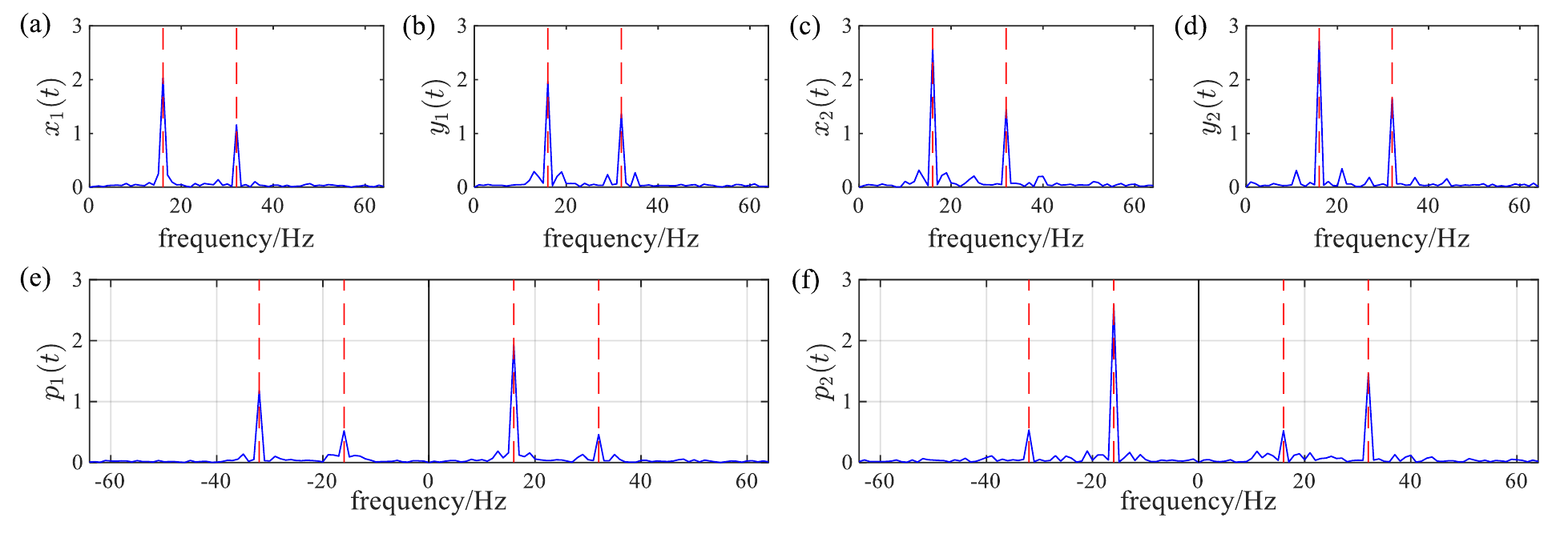}
    \caption{The spectrum and the full spectrum of the simulated signal. (a) and (b) Fourier spectrum of the real and imaginary parts for bearing section 1. (c) and (d) Fourier spectrum of the real and imaginary parts for bearing section 2. (e) and (f) Full spectrums of bearing section 1 and 2.}
    \label{figure3}
\end{figure}

Next, we constructed the complex-valued signals for the two bearing sections according to eq.(\ref{eq12}) in the manuscript as: 
\begin{equation}
    \left\{ \begin{array}{l}
        {p_1}(t) = {x_1}(t) + j{y_1}(t)\\
        {p_2}(t) = {x_2}(t) + j{y_2}(t)
        \end{array} \right.,
\end{equation}
 To be closer to the actual situation, the Gaussian noise with a signal-to-noise (SNR) ratio of 8.78 dB was added to the vibration response of the two bearing sections, respectively. The noisy signal waveforms and orbits are shown in Fig. \ref{figure2}. It can be seen that the amplitudes are time-varying, and the noise is strong. The Fourier transform spectrums and full spectrum are illustrated in Fig. \ref{figure3}. The full spectrum showed the information of the precession direction of different components, that is, the 1X component of the bearing section 1 was forward precession, whereas the 2X component was backward precession. The precession information of the bearing section 2 was completely opposite to that of the bearing section 1. The above analysis showed that the traditional method cannot present the non-stationary vibration process of the rotor system.
\begin{figure}
    \centering
    \includegraphics[width=0.8\linewidth]{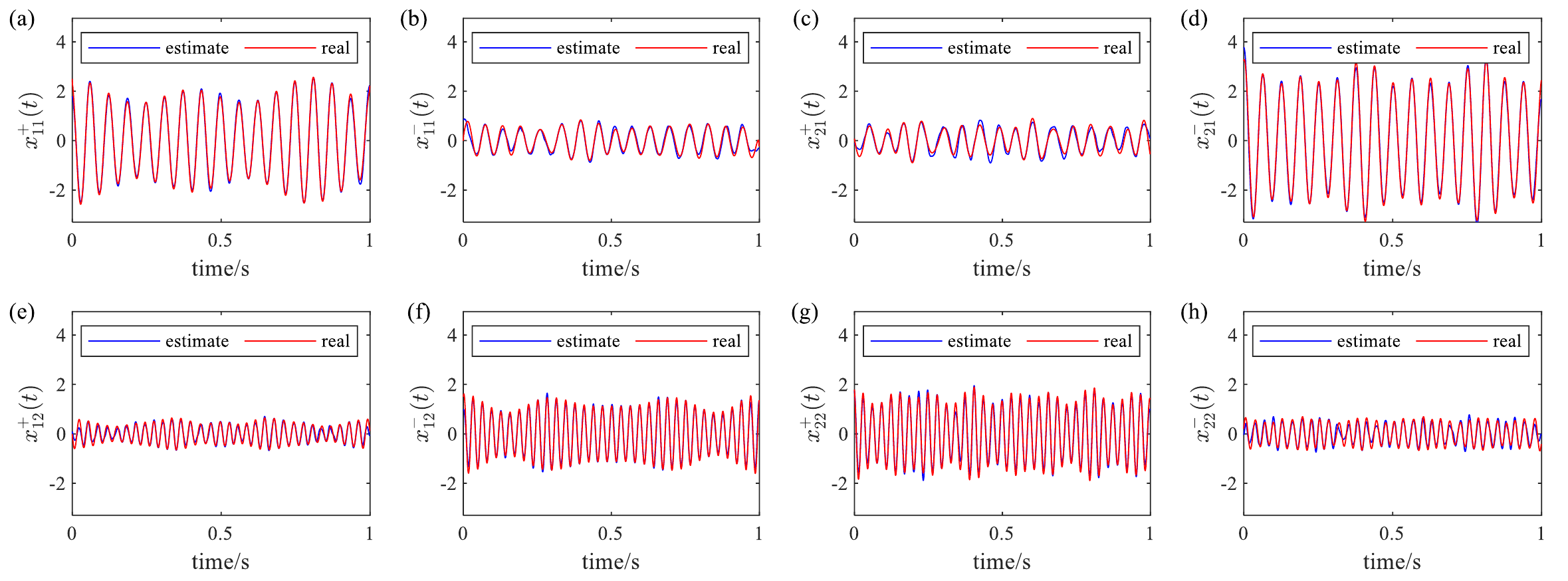}
    \caption{Decomposition results of the simulated signal using MVMD. (a) and (b) The forward and backward components of 1X component for bearing section 1. (c) and (d) The forward and backward components of 1X component for bearing section 2. (e) and (f) The forward and backward components of 2X component for bearing section 1. (g) and (h) The forward and backward components of 2X component for bearing section 2. (blue: estimate; red: real).}
    \label{figure4}
\end{figure}
\begin{figure}
    \centering
    \includegraphics[width=0.8\linewidth]{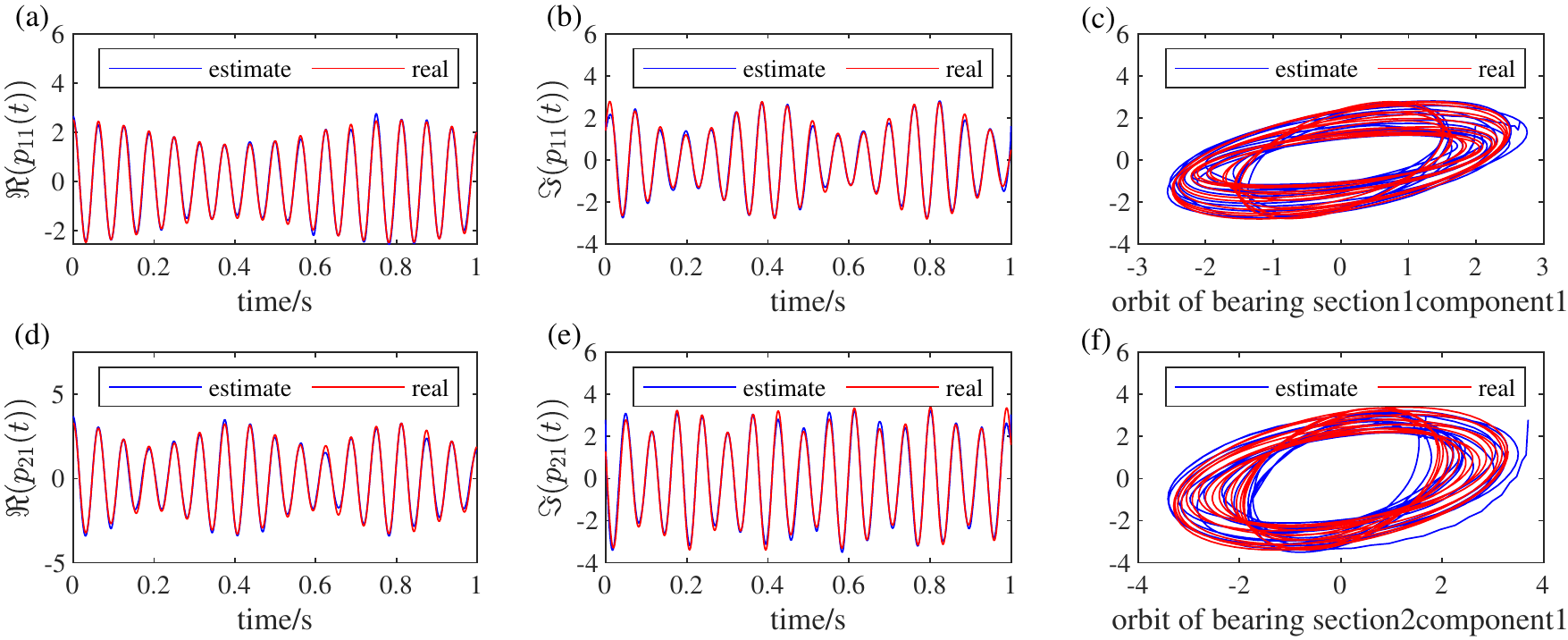}
    \caption{Signal reconstruction results (1X component). (a) and (b) Real and imaginary parts of bearing section 1. (c) Reconstructed orbit of bearing section 1. (d) and (e) Real and imaginary parts of bearing section 2. (f) Reconstructed orbit of bearing section 2. (blue: estimate; red: real).}
    \label{figure5}
\end{figure}
\begin{figure}
    \centering
    \includegraphics[width=0.8\linewidth]{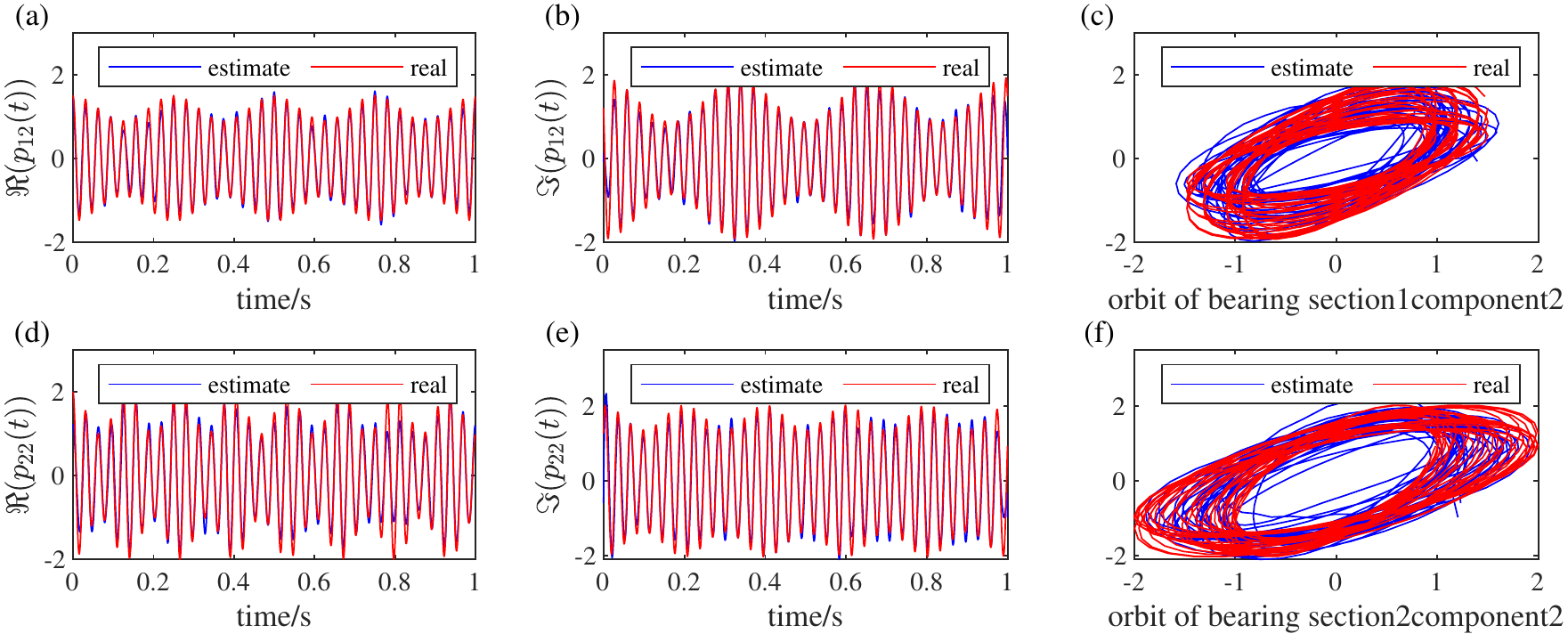}
    \caption{Signal reconstruction results (2X component). (a) and (b) Real and imaginary parts of bearing section 1. (c) Reconstructed orbit of bearing section 1. (d) and (e) Real and imaginary parts of bearing section 2. (f) Reconstructed orbit of bearing section 2. (blue: estimate; red: real).}
    \label{figure6}
\end{figure}
\begin{figure}
    \centering
    \includegraphics[width=0.8\linewidth]{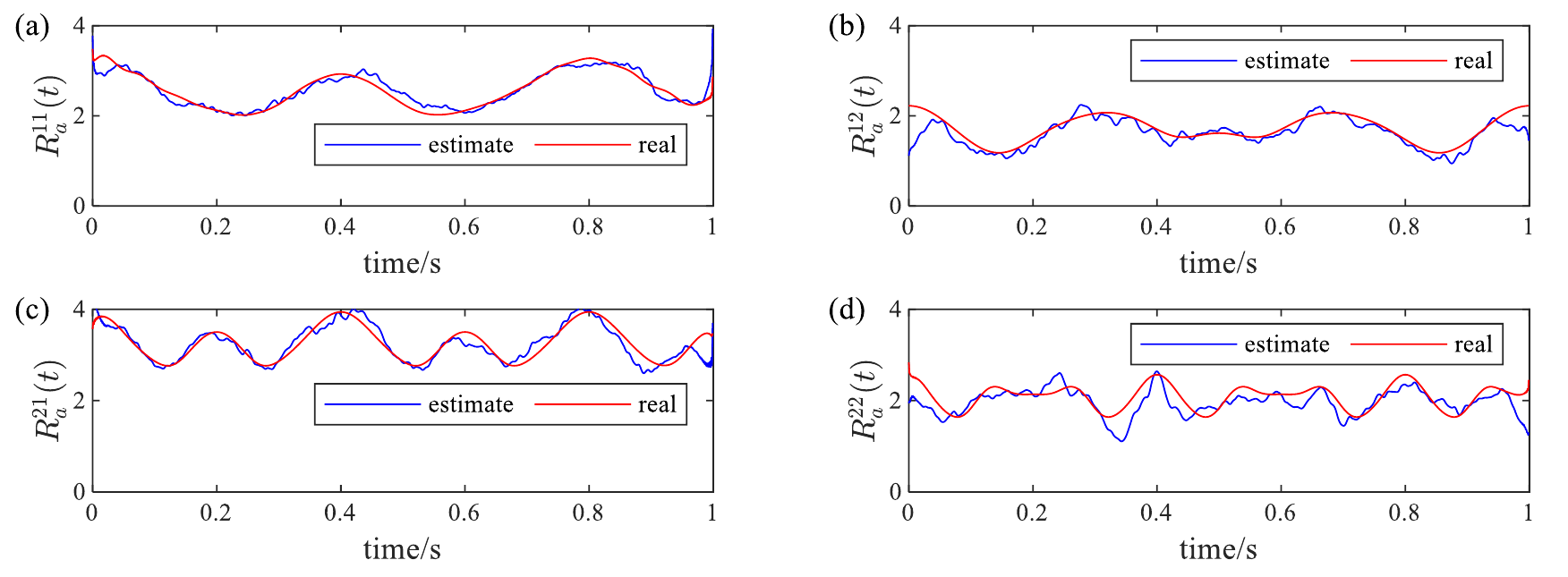}
    \caption{The semi-major axis. (a) and (b) The semi-major axis of 1X and 2X component for bearing section 1. (c) and (d) The semi-major axis of 1X and 2X component for bearing section 2. (blue: estimate; red: real).}
    \label{figure7}
\end{figure}
\begin{figure}
    \centering
    \includegraphics[width=0.8\linewidth]{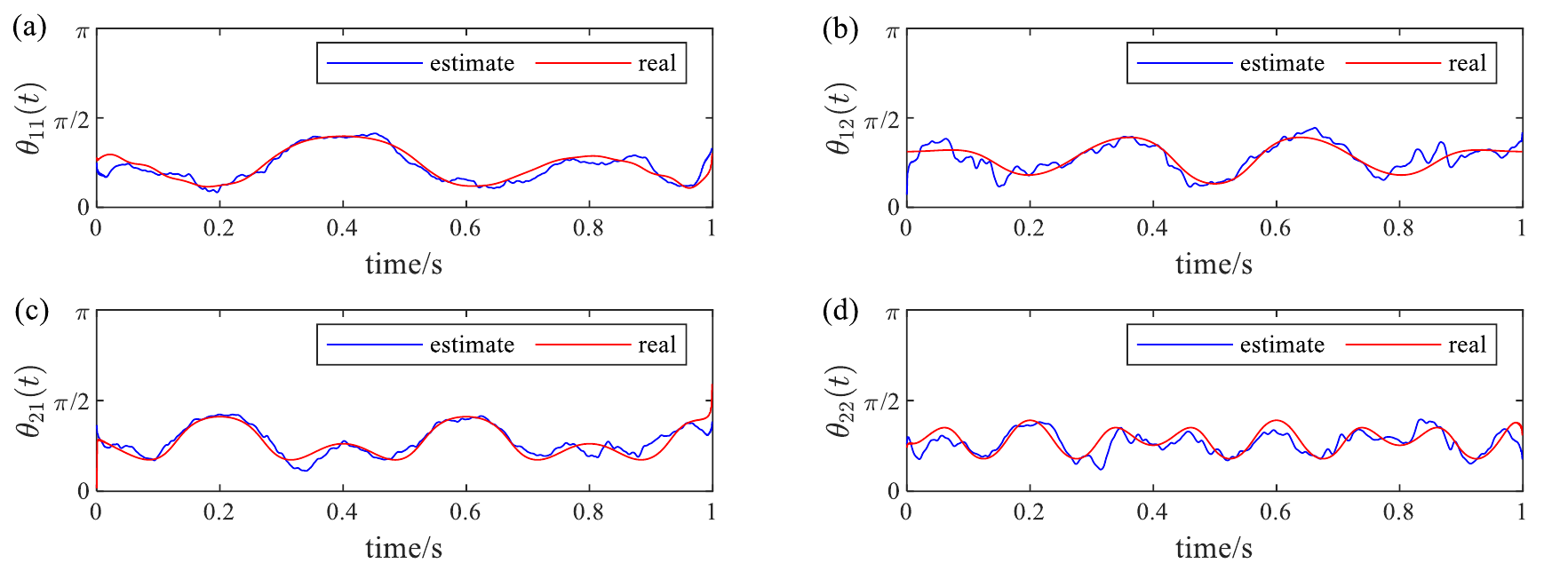}
    \caption{The instantaneous inclination angle. (a) and (b) The inclination angle of 1X and 2X component for bearing section 1. (c) and (d) The inclination angle of 1X and 2X component for bearing section 2. (blue: estimate; red: real).}
    \label{figure8}
\end{figure}
\begin{figure}
    \centering
    \includegraphics[width=0.8\linewidth]{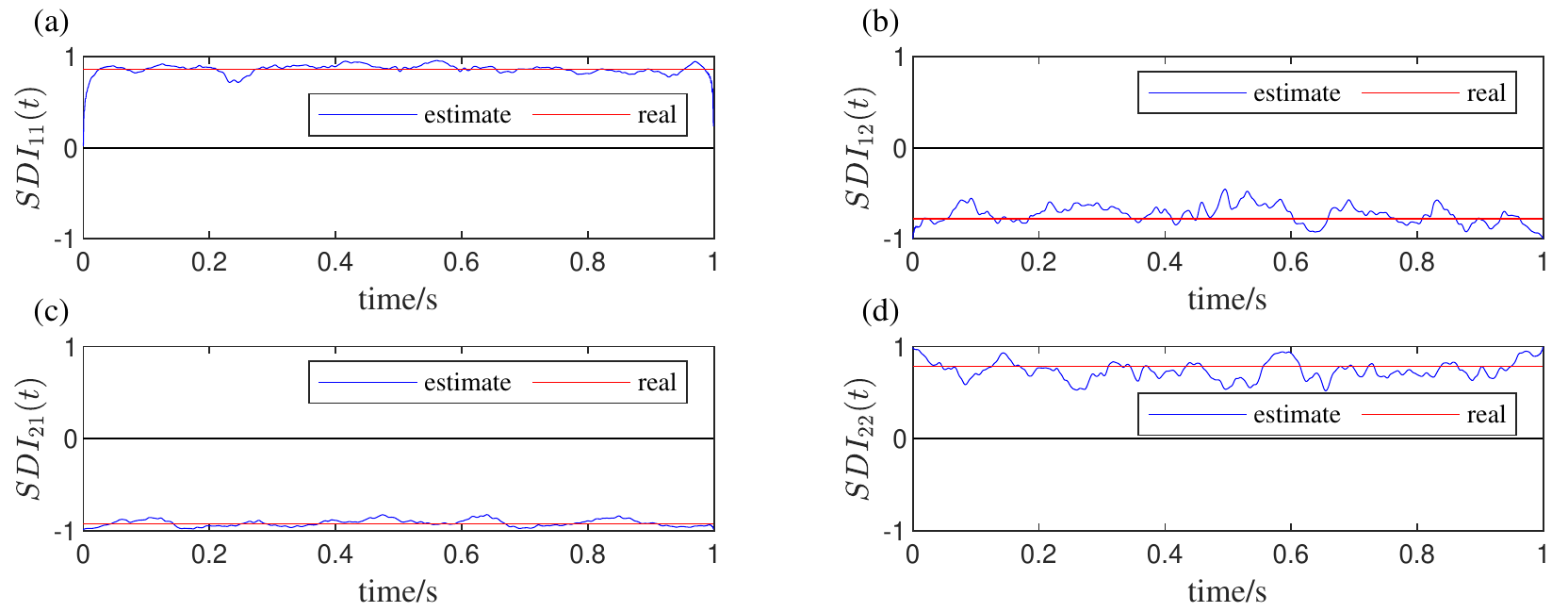}
    \caption{SDI. (a) and (b) SDI of 1X and 2X component for bearing section 1. (c) and (d) SDI of 1X and 2X component for bearing section 2. (blue: estimate; red: real).}
    \label{figure9}
\end{figure}

\begin{figure}
    \centering
    \includegraphics[width=1\linewidth]{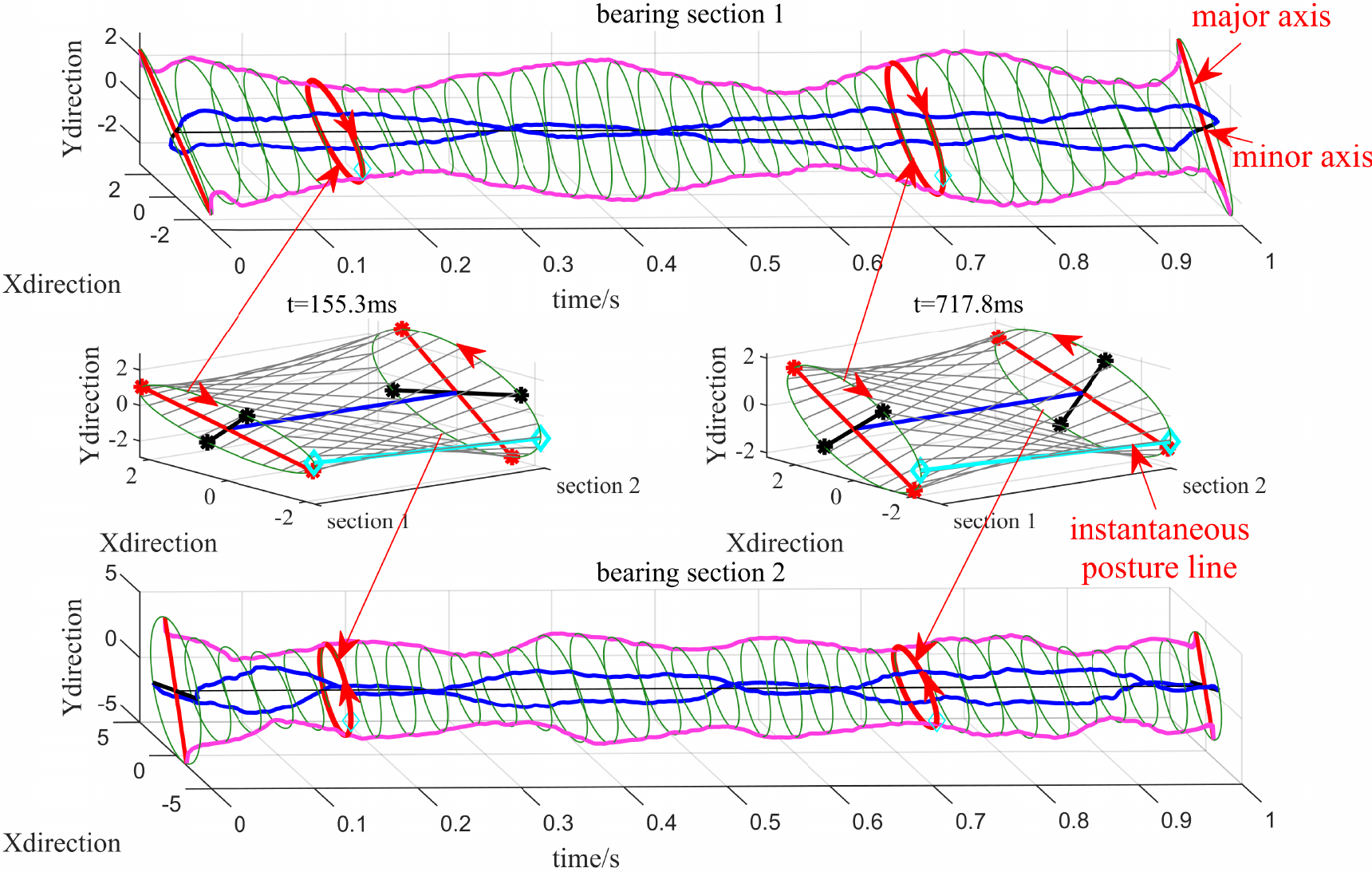}
    \caption{3D-IOM of the 1X component of the simulated signal for two bearing sections. }
    \label{figure10}
\end{figure}

\begin{figure}
    \centering
    \includegraphics[width=0.5\linewidth]{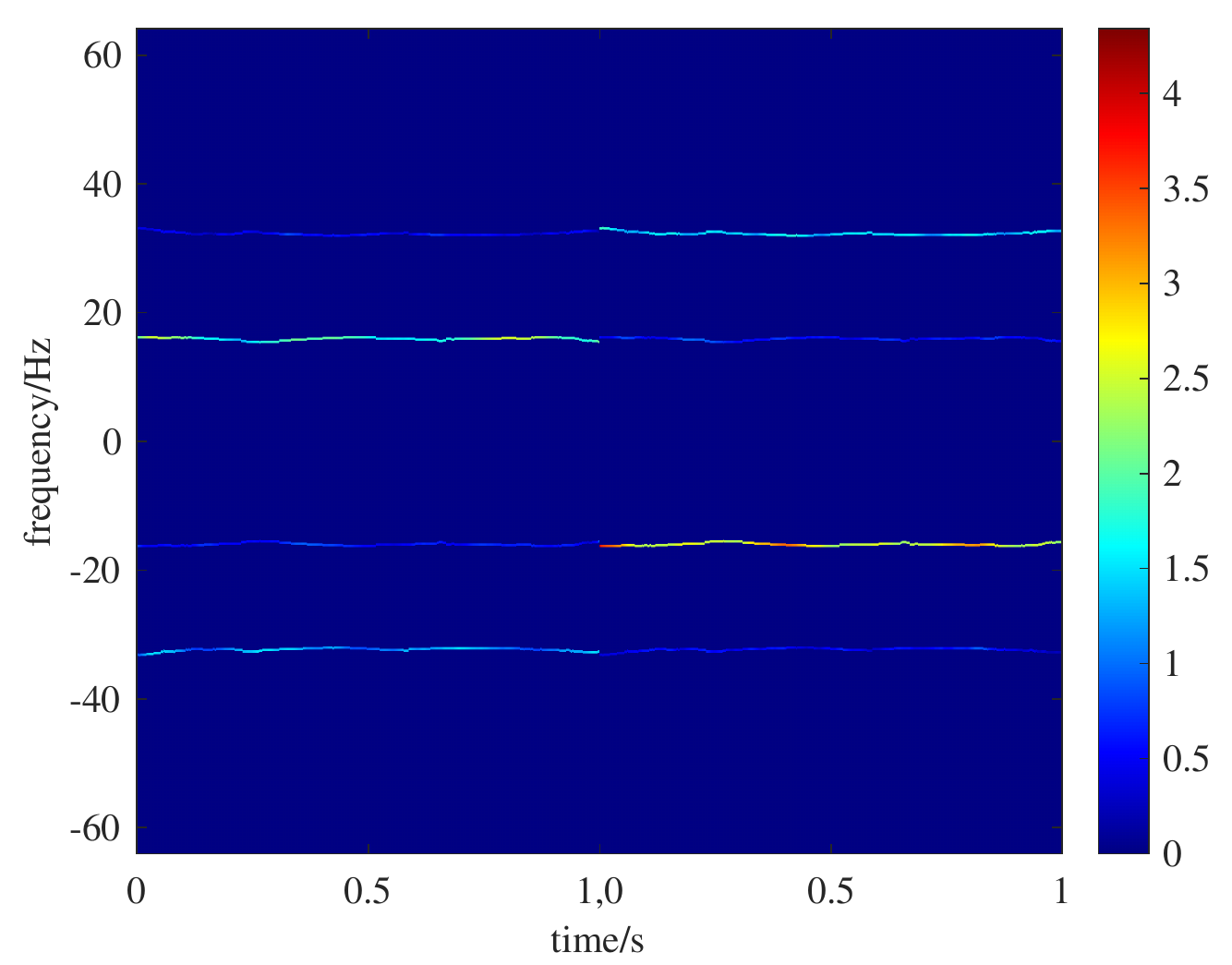}
    \caption{Time-FS of the simulated signal.}
    \label{figure11}
\end{figure}

A set of multi-component real-value signals $\{ {p_{1 + }}(t),{p_{1 - }}(t),{p_{2 + }}(t),{p_{2 - }}(t)\} $ could be obtained by eq. (\ref{eq20}) and which were then decomposed with MVMD. Fig. \ref{figure4} shows the decomposition results. It can be seen that the decomposition results were very close to the true values. The signals were separated according to the forward/backward frequency components, so the precession directions of the component on multiple rotor bearing sections could be determined by the amplitude of the signal waveform. The information of the precession direction obtained in Fig. \ref{figure3} and \ref{figure4} is consistent. In addition, we gave the reconstructed component signals and orbit (Fig. \ref{figure5} (1X components) and Fig. \ref{figure6} (2X components)). The results of reconstructed signals were well consistent with the real ones, and the orbits were more readable than the original ones. Next, the characteristic parameters of the instantaneous orbit are calculated based on the formula in section \ref{section3}. The semi-major axis of the instantaneous orbit, inclination angle, and SDI are displayed in Fig. \ref{figure7}, \ref{figure8} and \ref{figure9}, respectively. It can be seen from these figures that the instantaneous characteristic parameters of the orbit are time-varying and non-stationary and the proposed method achieved accurate estimation of each instantaneous parameter of the orbit. The 3D-IOM of the rotor is presented in Fig. \ref{figure10}. We drew the 3D-IOM of the rotor at t=155.3 and 717.8 ms. It can be observed from the 3D-IOM that the shape and instantaneous inclination angle of the orbit were changing slightly with time. The above analysis indicated that the instantaneous state of the rotor shaft could be determined at any time after the acquisition of instantaneous characteristics for the orbit. The time-FS of the simulated signal is depicted in Fig. \ref{figure11}. The relationship between the vibration intensity of the different components on the two bearing sections can be observed, and the information about the precession direction of each component on the bearing section.

\subsection{Analysis of the bistable behavior of roter system}
The bistable behavior of the rotor is a nonlinear behavior of the rotor-bearing system. The rotor jumps from the current stable state to another stable state after a certain excitation. However, the cause of the bistable behavior is for further study \cite{qu2007holospectrum}. In this study, a blower rotor with a working speed of 5500 rpm was selected to explore the bistable behavior of its vibration. The sampling frequency was set to 2000 Hz, as well as the number of sampling points was 1024.

\begin{figure}
    \centering
    \includegraphics[width=0.8\linewidth]{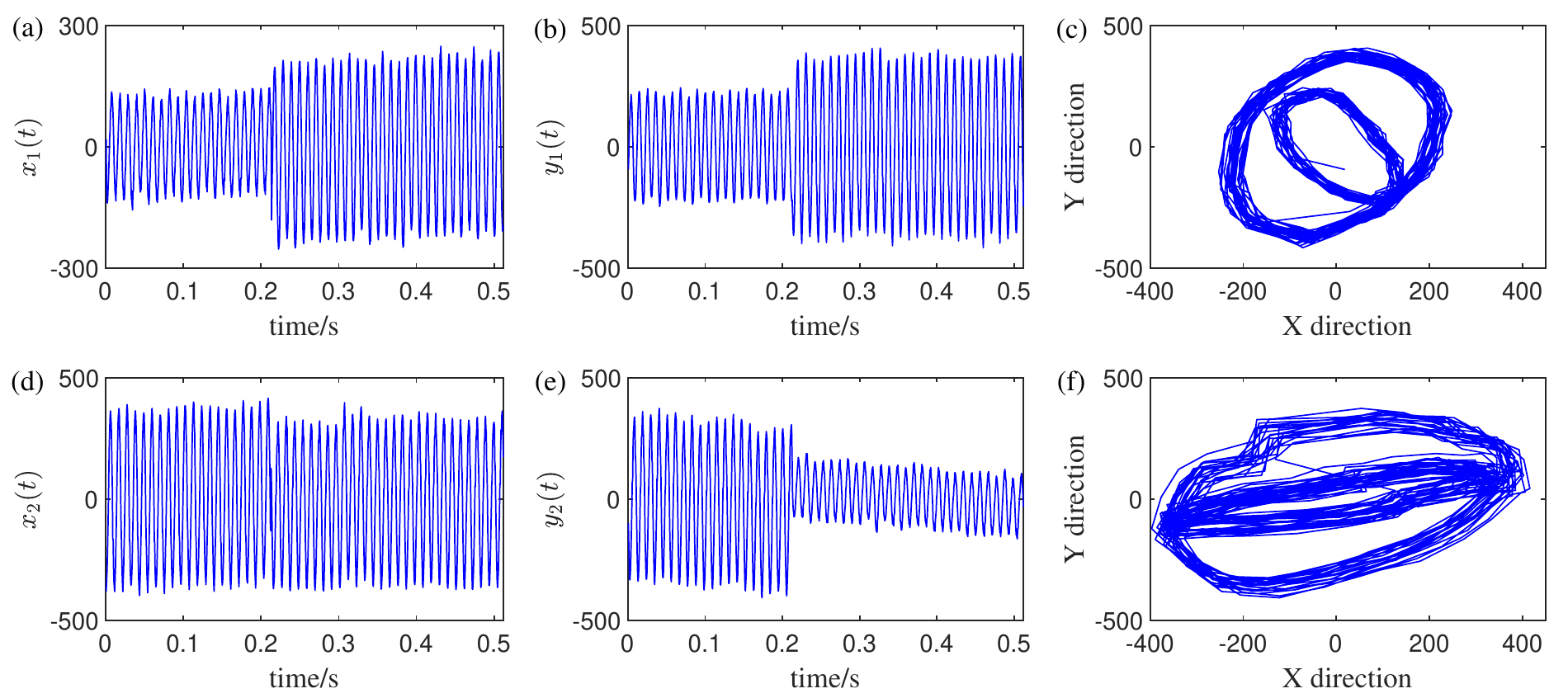}
    \caption{The signal waveform and orbit. (a) and (b) Real and imaginary parts of bearing section 1. (c) Orbit of bearing section 1. (d) and (e) Real and imaginary parts of bearing section 2. (f) Orbit of bearing section 2.}
    \label{figure12}
\end{figure}

\begin{figure}
    \centering
    \includegraphics[width=0.8\linewidth]{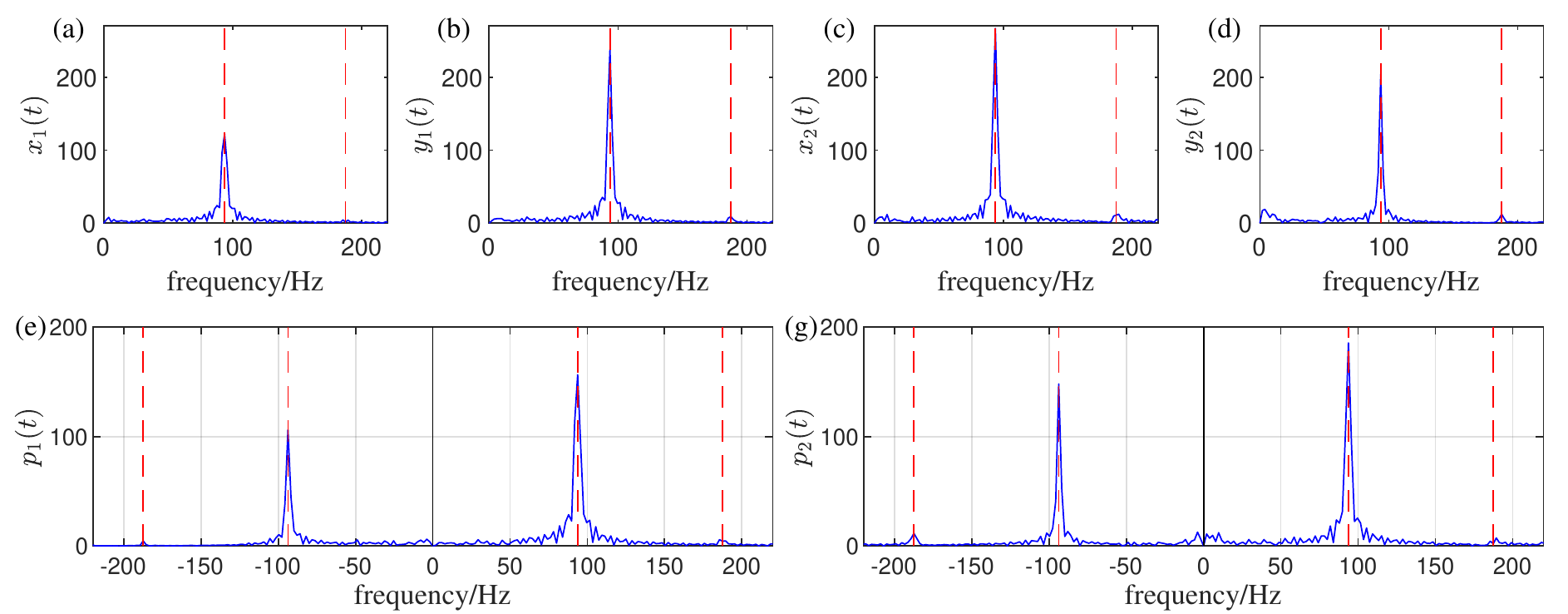}
    \caption{The Fourier spectrum and the full spectrum. (a) and (b) Fourier spectrum of the real and imaginary parts for bearing section 1. (c) and (d) Fourier spectrum of the real and imaginary parts for bearing section 2. (e) and (f) Full spectrums of bearing section 1 and 2.}
    \label{figure13}
\end{figure}

\begin{figure}
    \centering
    \includegraphics[width=0.8\linewidth]{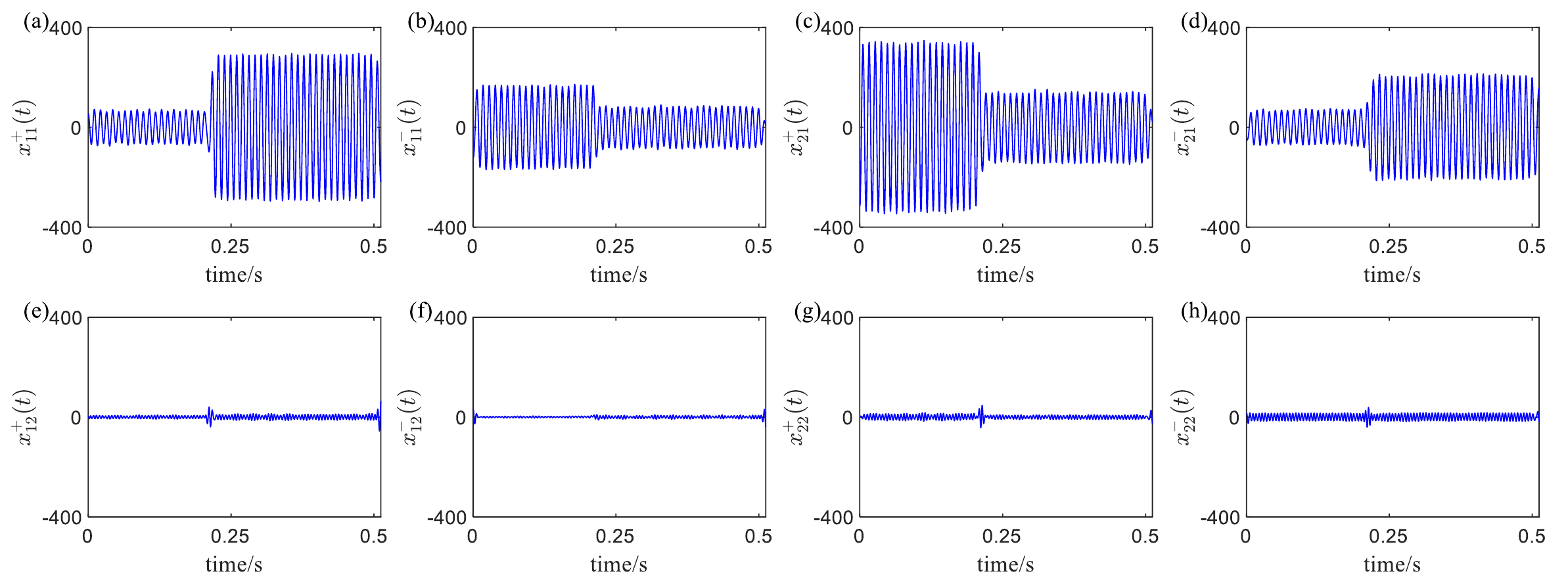}
    \caption{Decomposition results of bistable signal using MVMD. (a) and (b) The forward and backward components of 1X component for bearing section 1. (c) and (d) The forward and backward components of 1X component for bearing section 2. (e) and (f) The forward and backward components of 2X component for bearing section 1. (g) and (h) The forward and backward components of 2X component for bearing section 2.}
    \label{figure14}
\end{figure}

\begin{figure}
    \centering
    \includegraphics[width=0.8\linewidth]{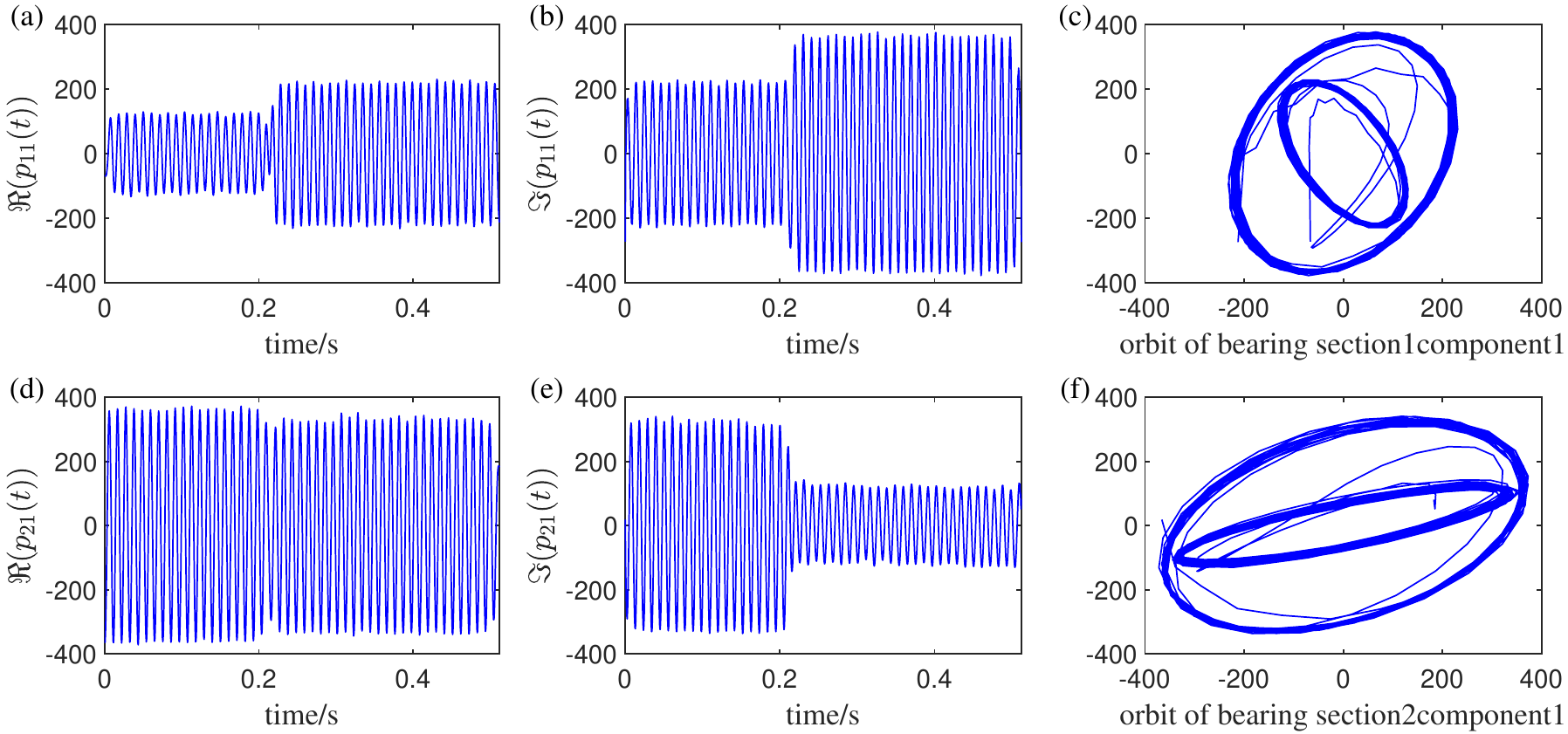}
    \caption{Reconstructed signal (1X component). (a) and (b) Real and imaginary parts of bearing section 1. (c) Reconstructed orbit of bearing section 1. (d) and (e) Real and imaginary parts of bearing section 2. (f) Reconstructed orbit of bearing section 2.}
    \label{figure15}
\end{figure}

The signal waveform and the orbit of the bearing sections at both ends of the rotor are shown in Fig. \ref{figure12}. It can be observed that the signal jumps around 0.2s. The Fourier spectrum and the full spectrum of the signal are depicted in Fig. \ref{figure13}. Due to the averaging effect of the Fourier transform, the transient information had been lost, so the moment of sudden change was not displayed. The decomposition results of the bistable signal are shown in Fig. \ref{figure14}. The amplitude of the 2X component is very small. It is meaningless to discuss this situation. Therefore, we only discussed the 1X component of the bistable signal. Fig. \ref{figure15} shows the reconstructed signal of 1X component. Compared with the original orbit, the reconstructed component orbit became smoother.

\begin{figure}
    \centering
    \includegraphics[width=0.8\linewidth]{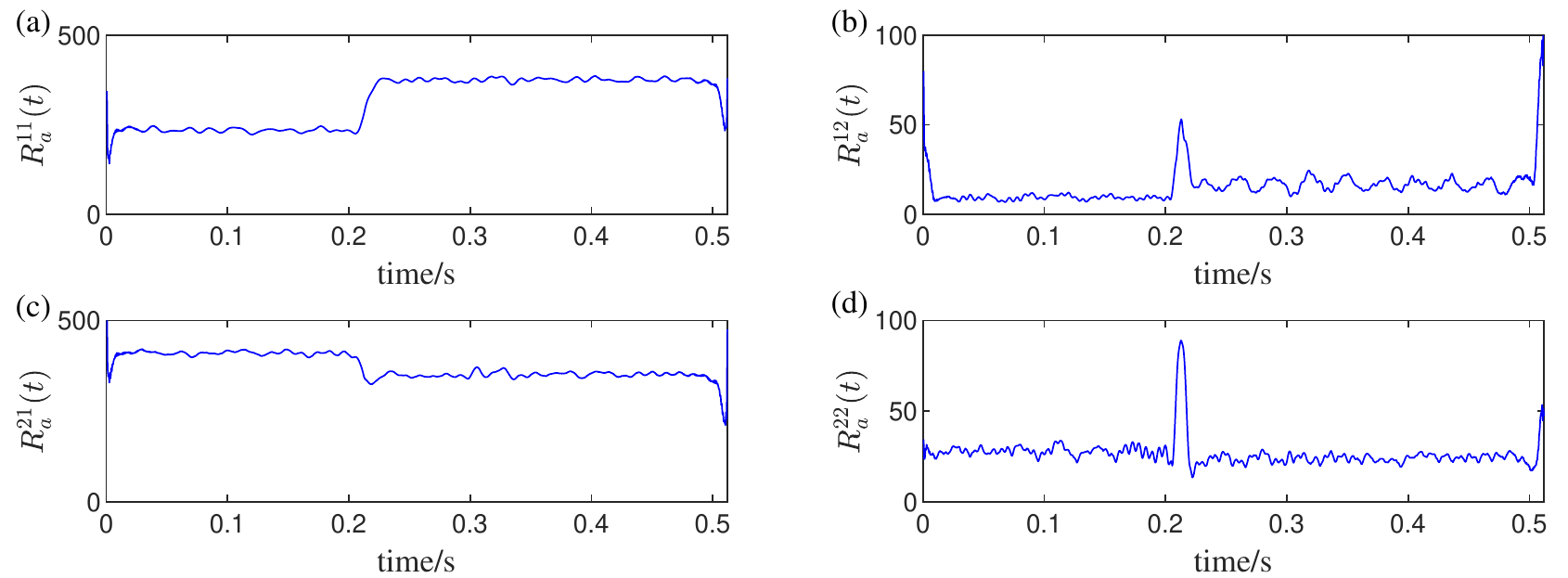}
    \caption{Semi-major axis. (a) and (b) The semi-major axis of 1X and 2X component for bearing section 1. (c) and (d) The semi-major axis of 1X and 2X component for bearing section 2.}
    \label{figure17}
\end{figure}

\begin{figure}
    \centering
    \includegraphics[width=0.8\linewidth]{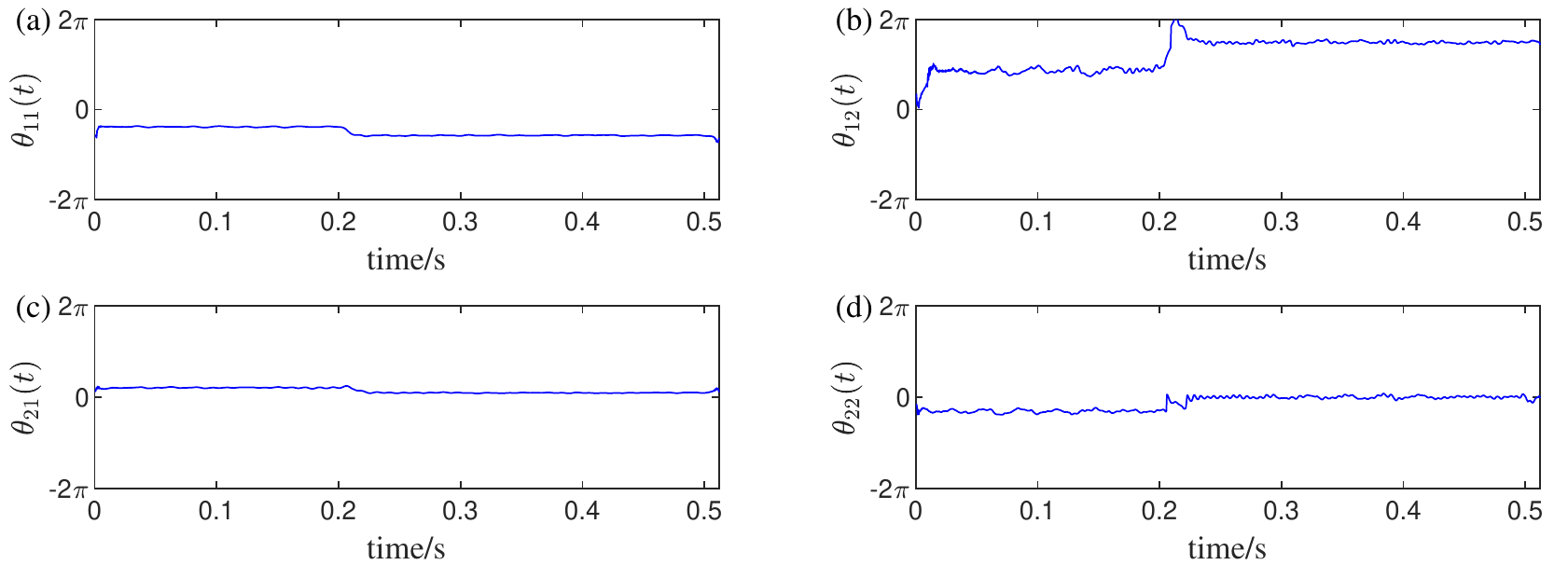}
    \caption{The instantaneous inclination angle. (a) and (b) The inclination angle of 1X and 2X component for bearing section 1. (c) and (d) The inclination angle of 1X and 2X component for bearing section 2.}
    \label{figure18}
\end{figure}

\begin{figure}
    \centering
    \includegraphics[width=0.8\linewidth]{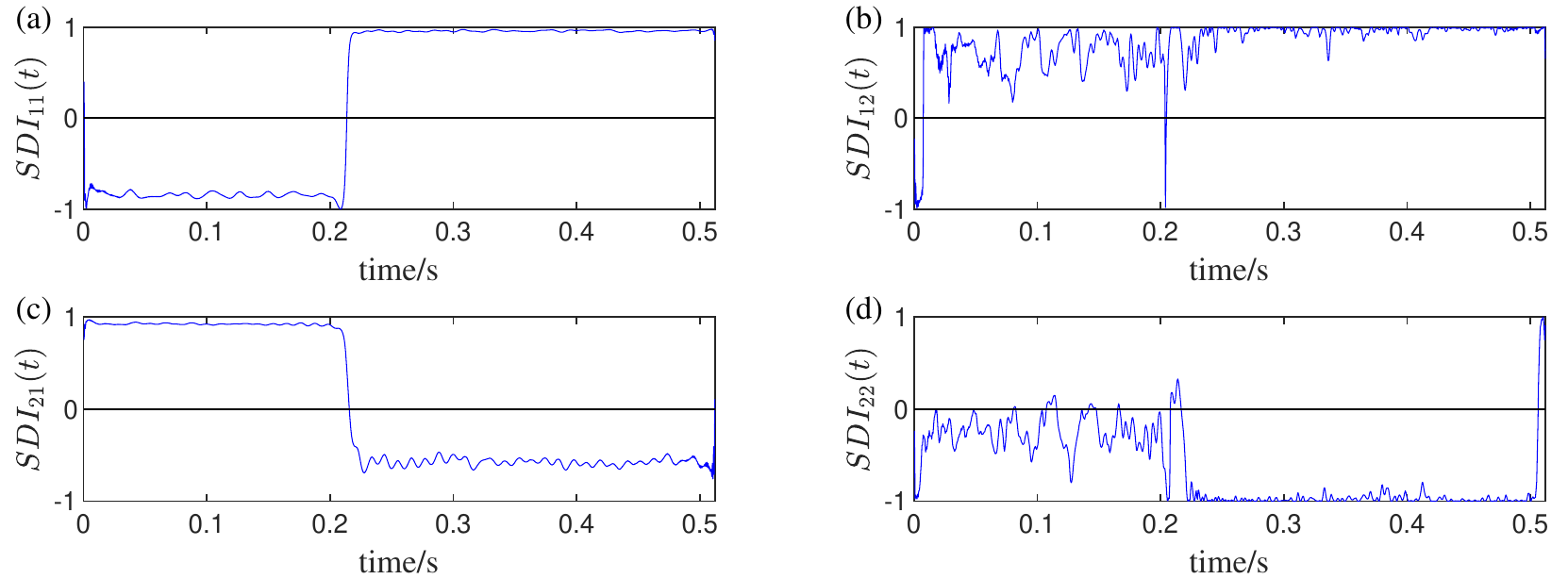}
    \caption{SDI. (a) and (b) SDI of 1X and 2X component for bearing section 1. (c) and (d) SDI of 1X and 2X component for bearing section 2.}
    \label{figure19}
\end{figure}

\begin{figure}
    \centering
    \includegraphics[width=1.0\linewidth]{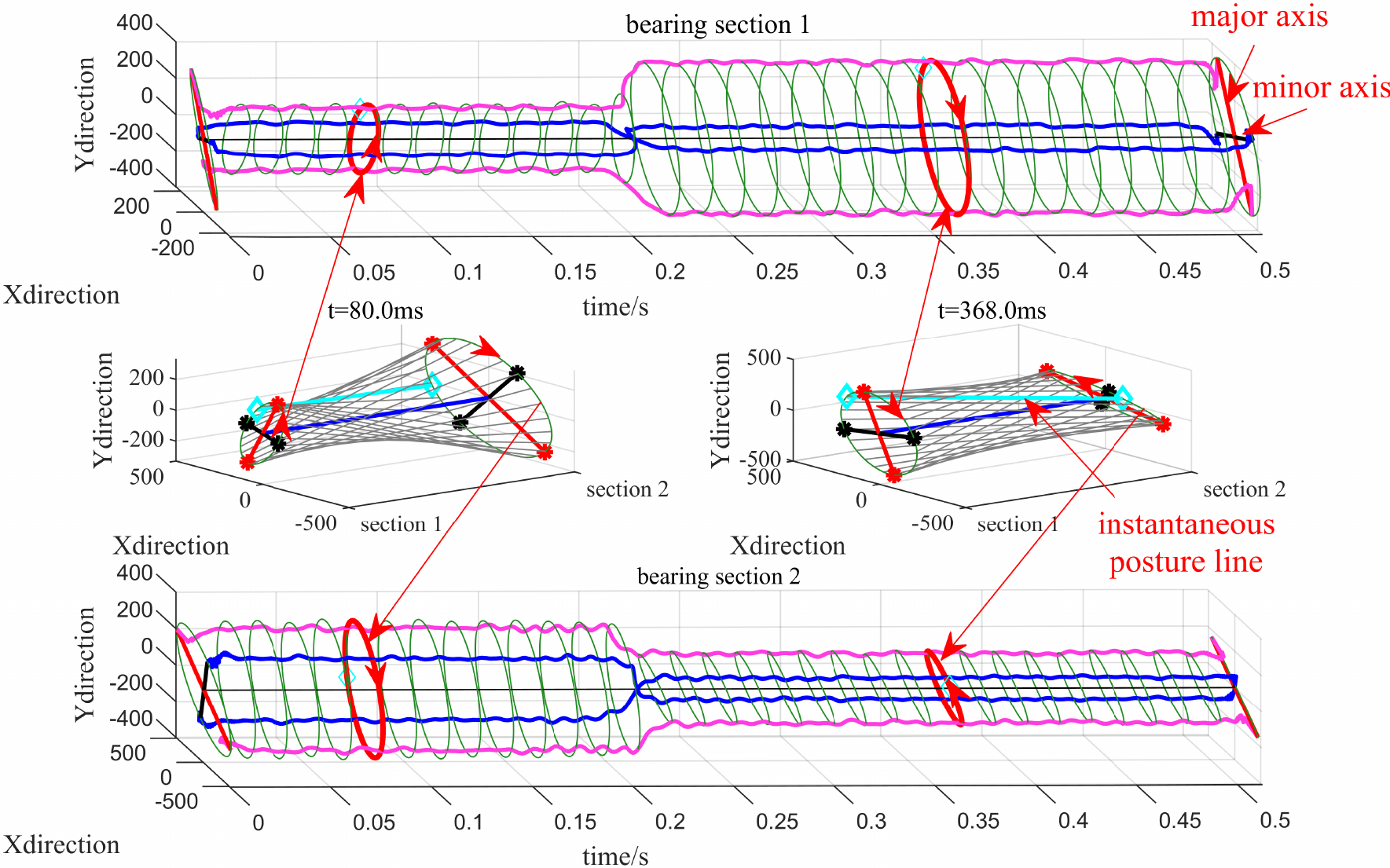}
    \caption{3D-IOM of the 1X component of the bistable signal on the two bearing sections.}
    \label{figure20}
\end{figure}

\begin{figure}
    \centering
    \includegraphics[width=0.5\linewidth]{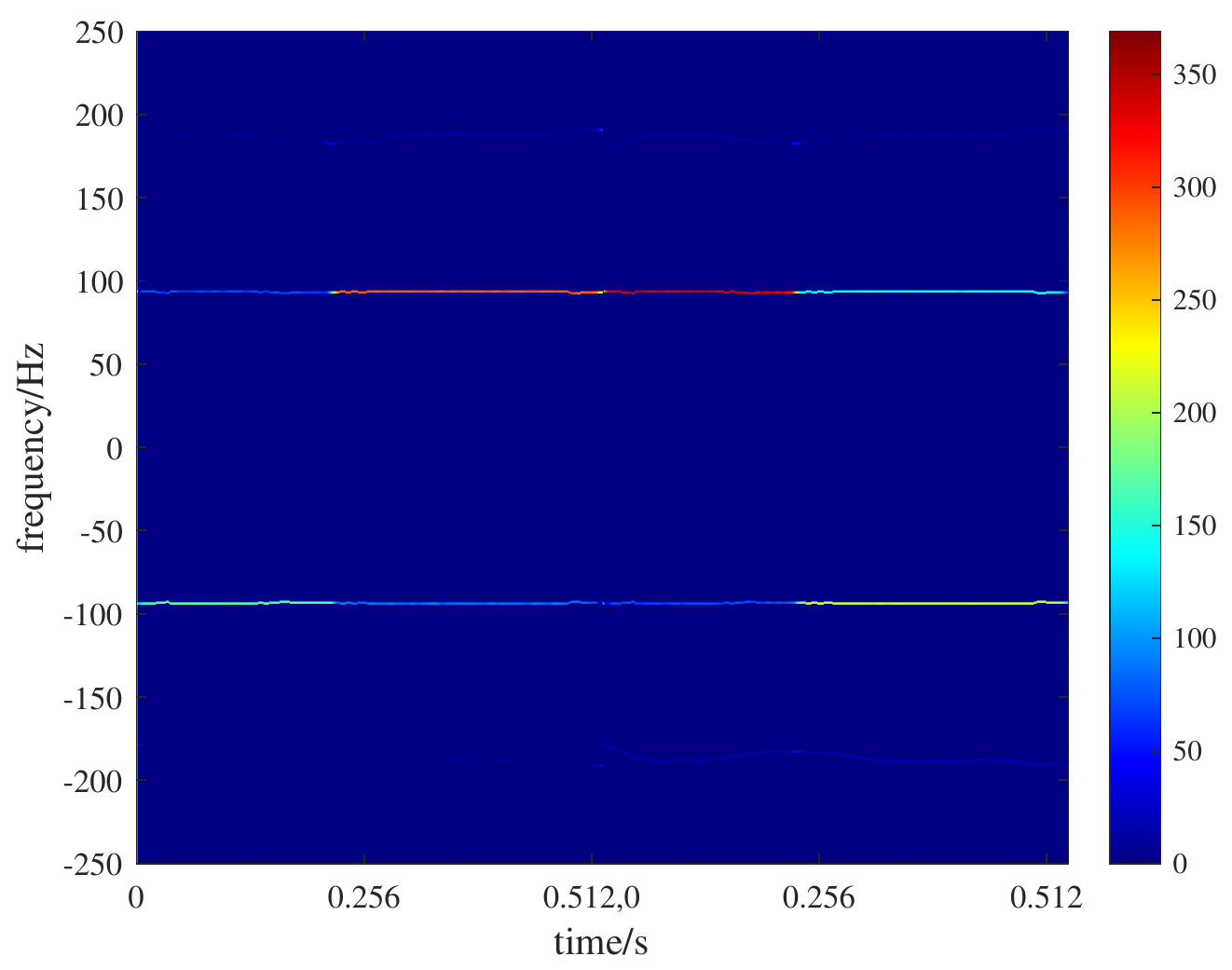}
    \caption{Time-FS of the bistable signal.}
    \label{figure21}
\end{figure}

The instantaneous parameters of each component orbit including the semi-major axis (Fig. \ref{figure17}), instantaneous inclination angle (Fig. \ref{figure18}), and SDI (Fig. \ref{figure19}) were obtained according to the construction method of 3D-IOM (Section \ref{section3}). The semi-major axis of the 1X component on the two bearing sections (Fig. \ref{figure17} (a) and (c)) jumped at 0.2s. This phenomenon was consistent with the results of the signal waveform. But Fig. \ref{figure17} shows the time-varying process of the signal. There was almost no change in the instantaneous inclination angle (Fig. \ref{figure18}). Figure \ref{figure19} highlighted the information that the SDI value of the 1X component had a zero-crossing jump around 0.2s, which means that the precession direction changed. The 3D-IOM is depicted in Fig. \ref{figure20}. The instantaneous inclination angle did not change significantly. It can be observed from the blue line that the precession directions of the 1X component on the two bearing sections are opposite before and after 0.2s. However, traditional methods cannot obtain this information. The Time-FS of the bistable signal is shown in Fig. \ref{figure21}. Corresponding with the spectrum, the frequencies of the forward/backward components of the two bearing sections were 96 Hz.

\subsection{Analysis of rotor vibration signals of a pumped storage unit}
\begin{figure}
    \centering
    \includegraphics[width=7cm,height=9cm]{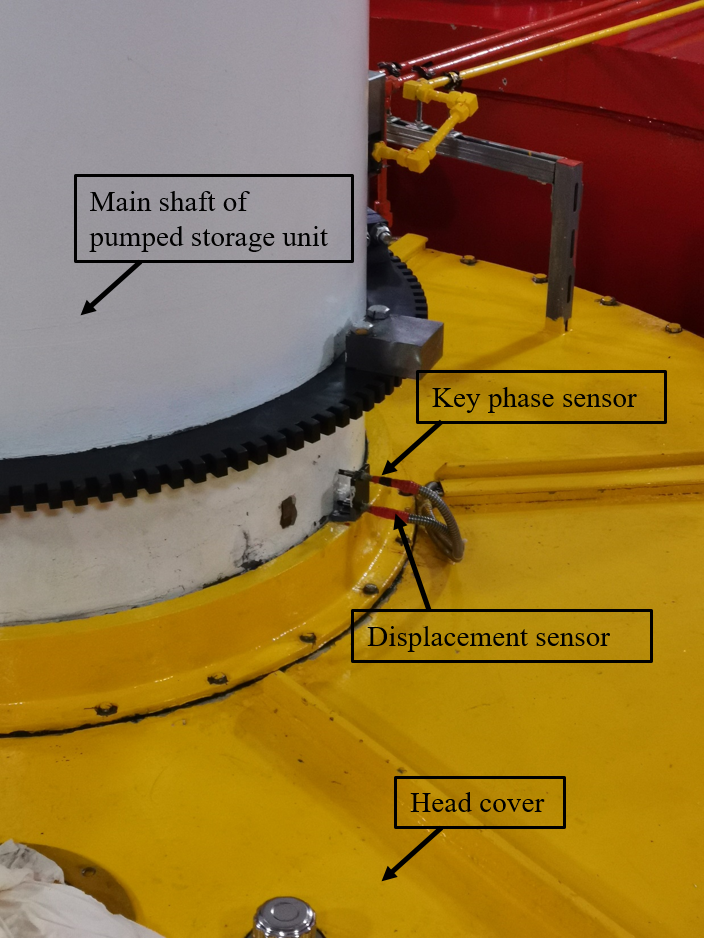}
    \caption{Displacement sensor for measuring vibration data of water guide bearing and key phase sensor}
    \label{figure33}
\end{figure}

The main parameters to characterize the operation and stability of the turbine are vibration, main shaft runout, and pressure pulsation, among which vibration is the main factor. Generally, most of the vibration factors are closely connected with shaft vibration. However, it is not enough to obtain the signal of only a single bearing section to analyze the fault of shaft vibration. The signal of one bearing section can only reflect the instantaneous characteristics of the shaft vibration process to a very limited extent and cannot reflect the non-stationary vibration process of the shaft as a whole. Therefore, it is necessary to analyze the vibration signals of multiple bearing sections simultaneously to obtain more meaningful and comprehensive joint information. In general, displacement sensors are arranged in mutually perpendicular directions of the upper guide bearing (UGB), lower guide bearing (LGB), and water guide bearing (WGB) of the hydraulic turbine to collect signals. As shown in Fig. \ref{figure33}, vibration data of the WGB is obtained by the displacement sensor. 

This study selected a set of shaft vibration data recorded by a vibration monitoring system for a pumped storage unit. The pumped storage unit had a rated speed of 375 rpm and a rated power of 400 MW. The sampling frequency was set to 800 Hz and the sampling time was 1.28 s, as well as the number of samples, is 1024.
\begin{figure}
    \centering
    \includegraphics[width=0.8\linewidth]{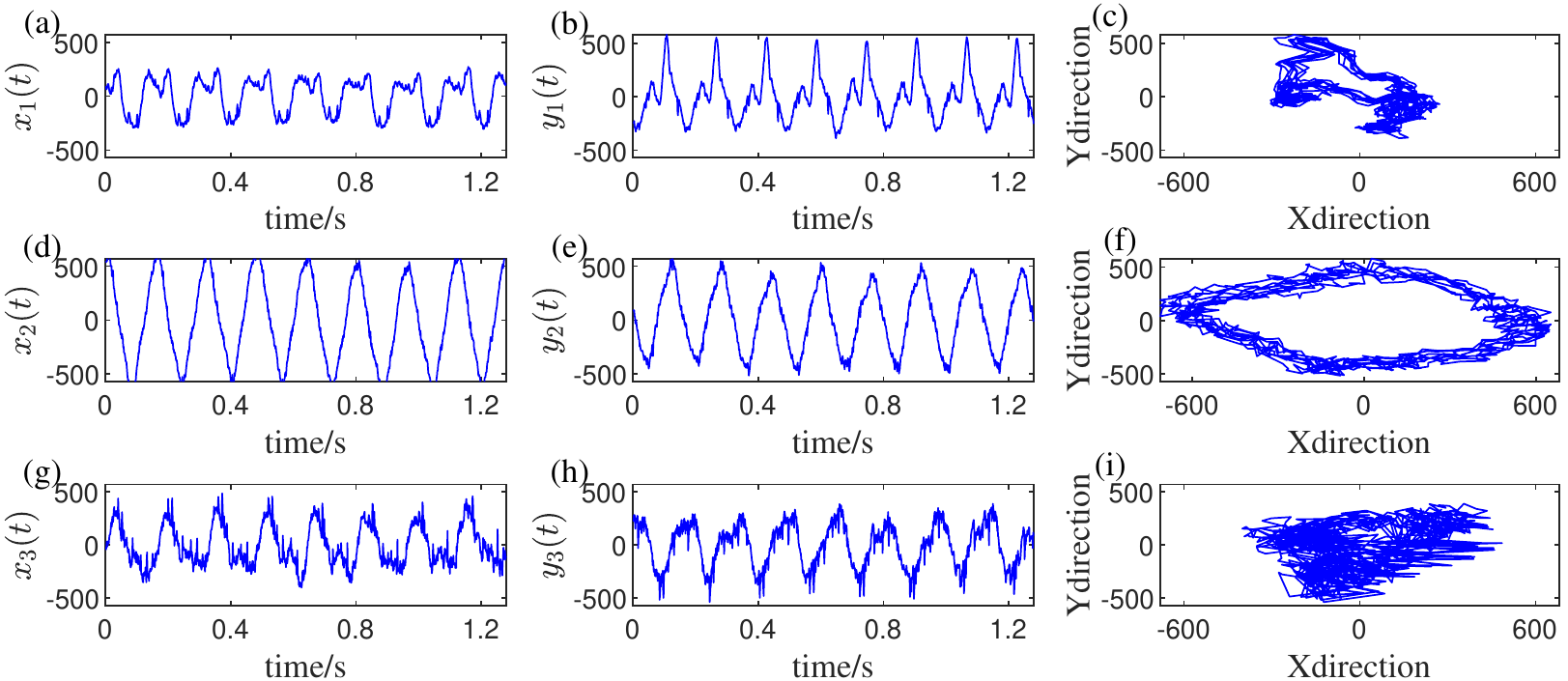}
    \caption{The noisy signal waveform and orbit. (a) and (b) Real and imaginary parts of the UGB. (c) Orbit of the UGB. (d) and (e) Real and imaginary parts of the LGB. (f) Orbit of the UGB. (g) and (h) Real and imaginary parts of the WGB. (i) Orbit of the WGB.}
    \label{figure22}
\end{figure}
\begin{figure}
    \centering
    \includegraphics[width=0.8\linewidth]{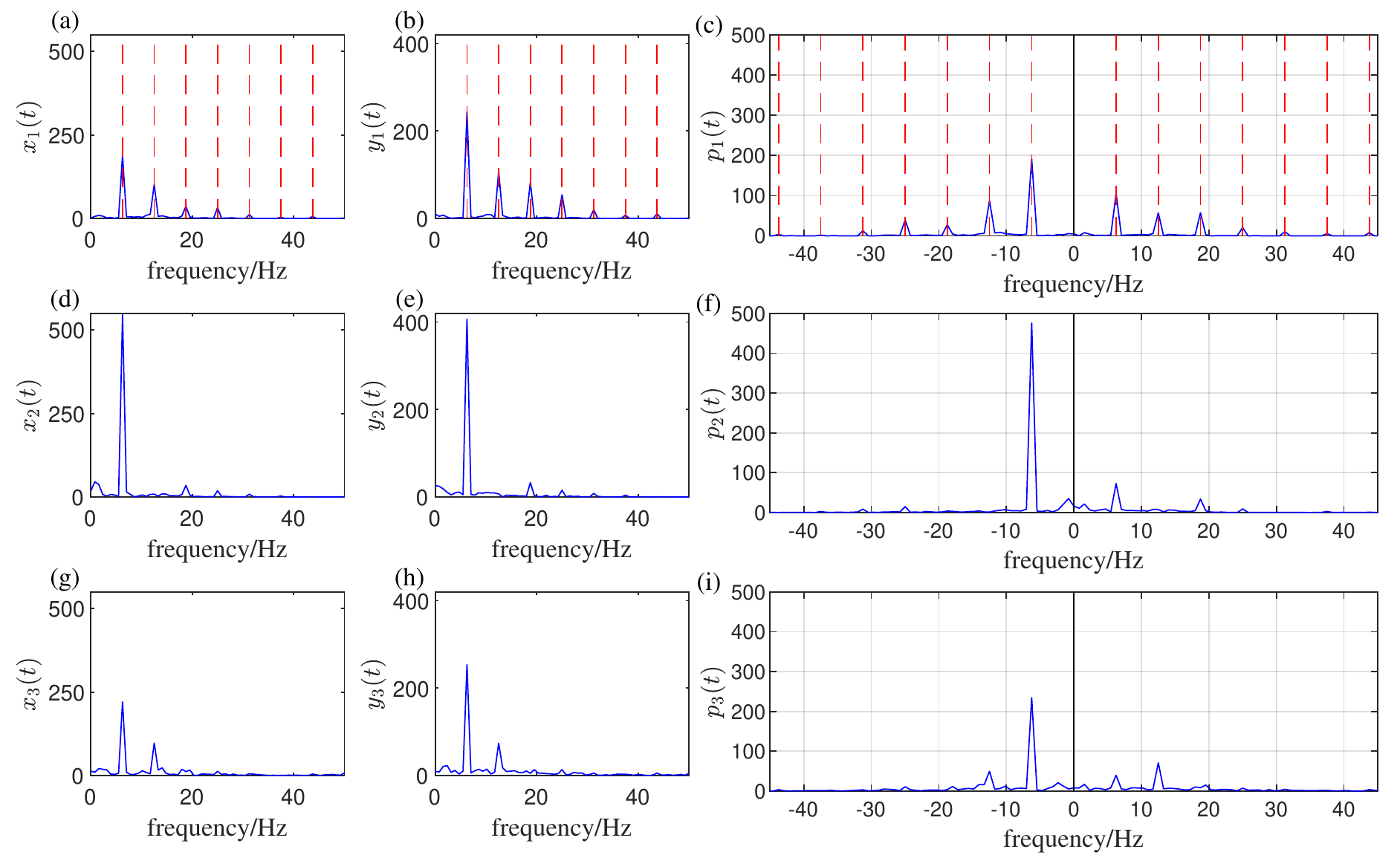}
    \caption{The spectrum and the full spectrum of the vibration signals. (a) and (b) Fourier spectrum of the real and imaginary parts for UGB. (d) and (e) Fourier spectrum of the real and imaginary parts for LGB. (g) and (h) Fourier spectrum of the real and imaginary parts for WGB. (c) (f) and (i) Full spectrum of the UGB, the LGB, and the WGB, respectively.}
    \label{figure23}
\end{figure}
\begin{figure}
    \centering
    \includegraphics[width=0.8\linewidth]{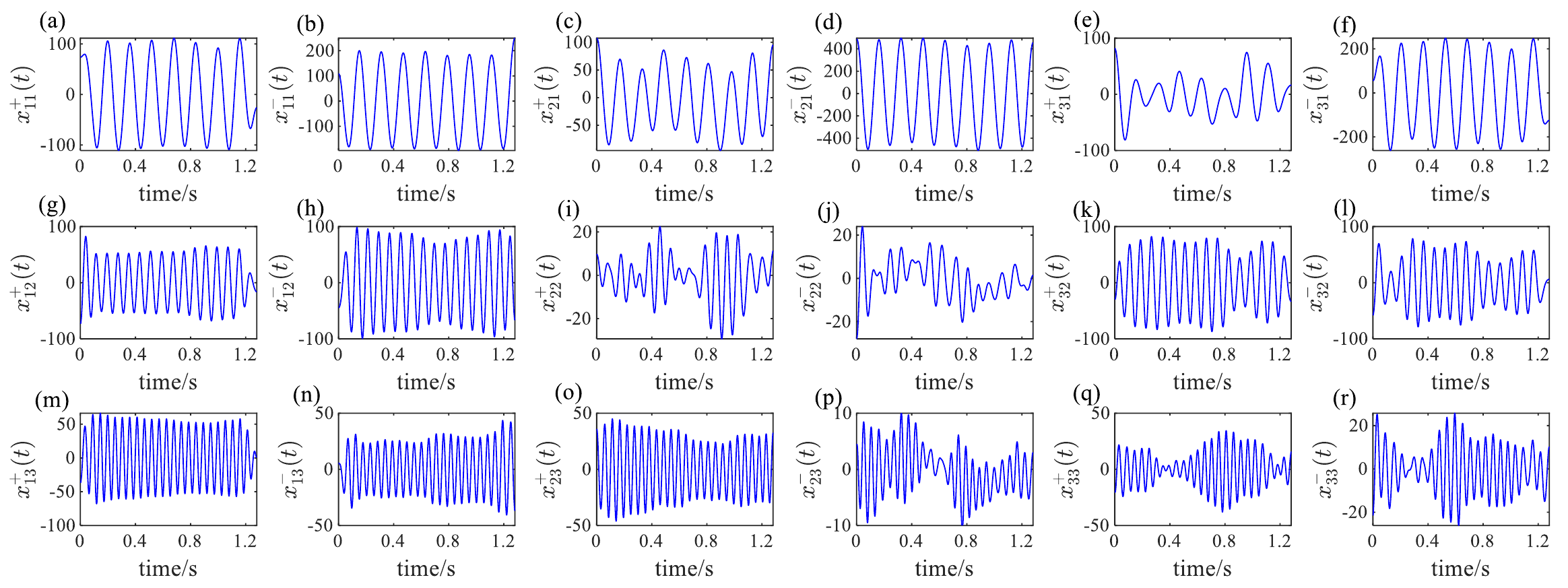}
    \caption{Decomposition results of rotor vibration signal using MVMD. (a)-(f) The forward and backward components of the 1X component for the UGB, LGB, and WGB, respectively. (g)-(l) The forward and backward components of the 2X component for the UGB, LGB, and WGB, respectively. (m)-(r) The forward and backward components of the 3X component for the UGB, LGB, and WGB, respectively.}
    \label{figure24}
\end{figure}
\begin{figure}
    \centering
    \includegraphics[width=0.8\linewidth]{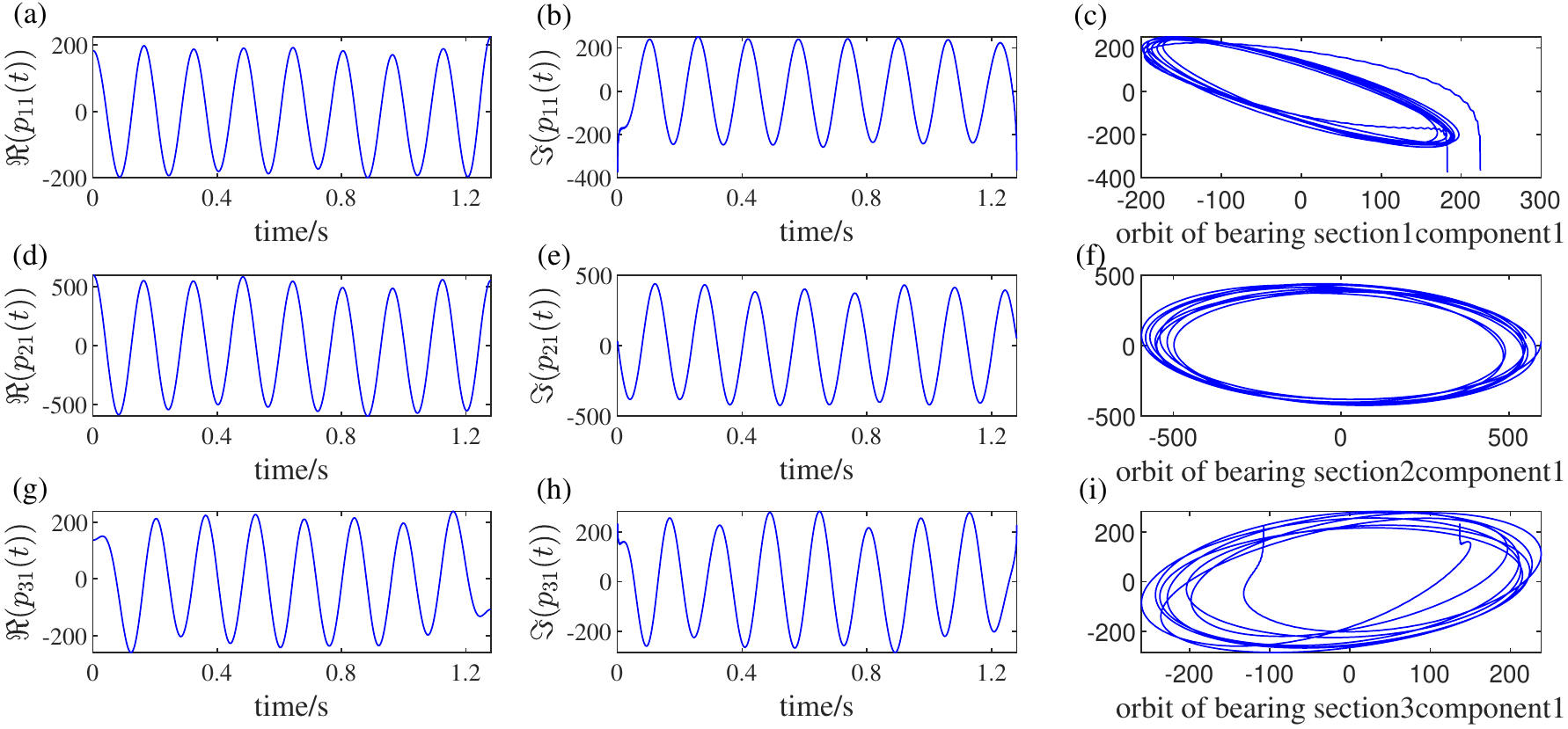}
    \caption{Reconstructed signal (1X components). (a) and (b) Real and imaginary parts of UGB. (c) Reconstructed orbit of UGB. (d) and (e) Real and imaginary parts of LGB. (f) Reconstructed orbit of LGB. (g) and (h) Real and imaginary parts of WGB. (i) Reconstructed orbit of WGB.}
    \label{figure25}
\end{figure}
\begin{figure}
    \centering
    \includegraphics[width=0.8\linewidth]{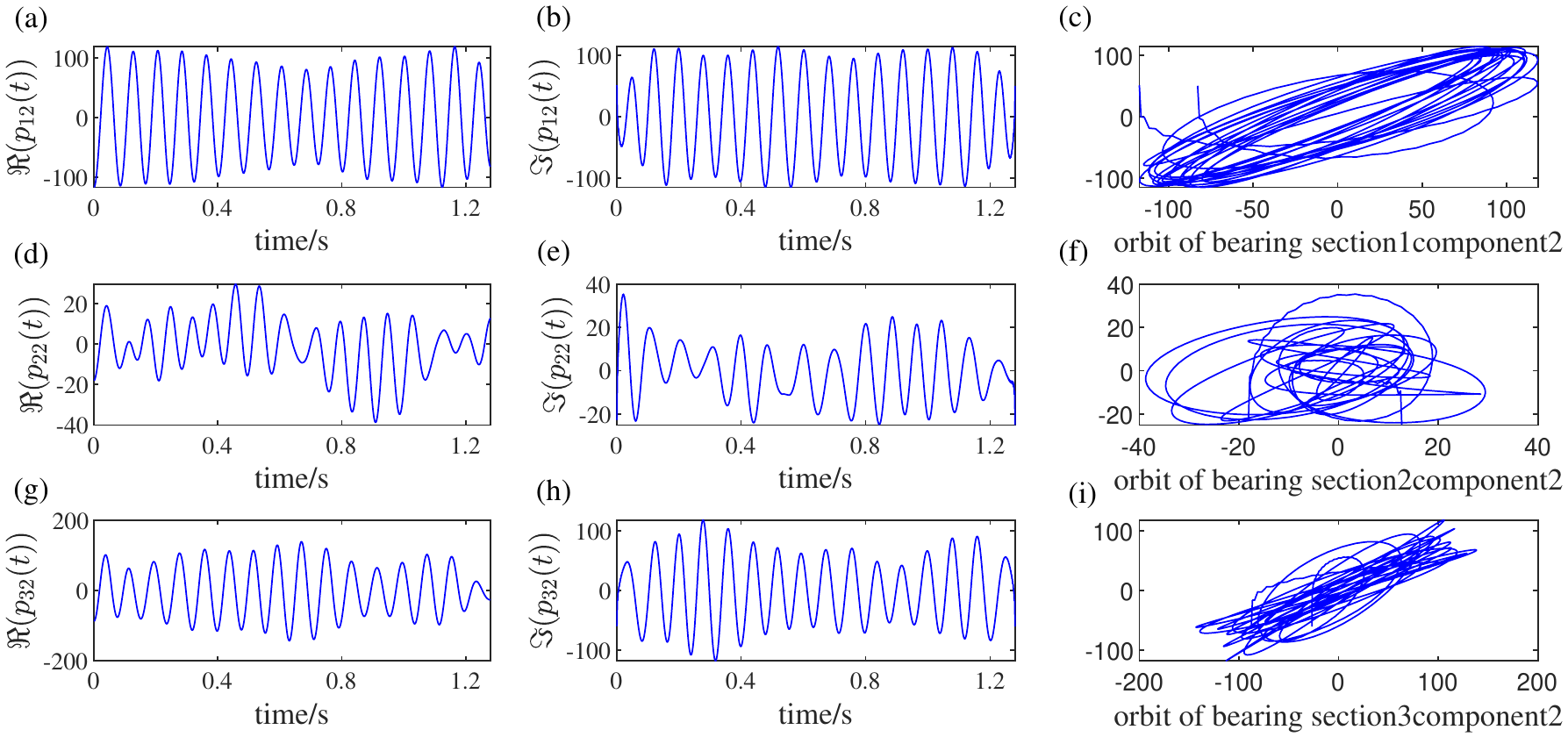}
    \caption{Reconstructed signal (2X components). (a) and (b) Real and imaginary parts of UGB. (c) Reconstructed orbit of UGB. (d) and (e) Real and imaginary parts of LGB. (f) Reconstructed orbit of LGB. (g) and (h) Real and imaginary parts of WGB. (i) Reconstructed orbit of WGB.}
    \label{figure26}
\end{figure}
\begin{figure}
    \centering
    \includegraphics[width=0.8\linewidth]{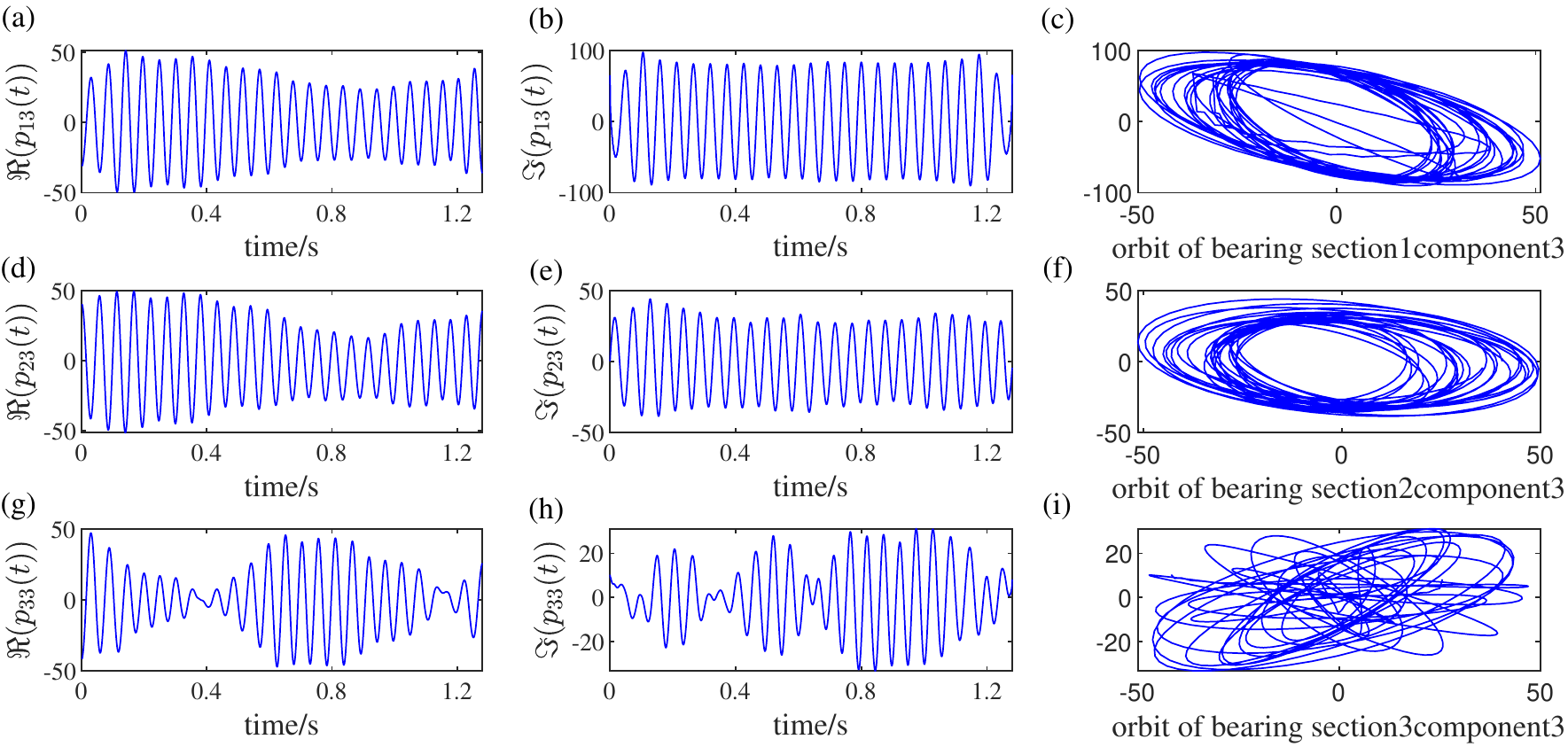}
    \caption{Reconstructed signal (3X components). (a) and (b) Real and imaginary parts of UGB. (c) Reconstructed orbit of UGB. (d) and (e) Real and imaginary parts of LGB. (f) Reconstructed orbit of LGB. (g) and (h) Real and imaginary parts of WGB. (i) Reconstructed orbit of WGB.}
    \label{figure27}
\end{figure}
\begin{figure}
    \centering
    \includegraphics[width=0.8\linewidth]{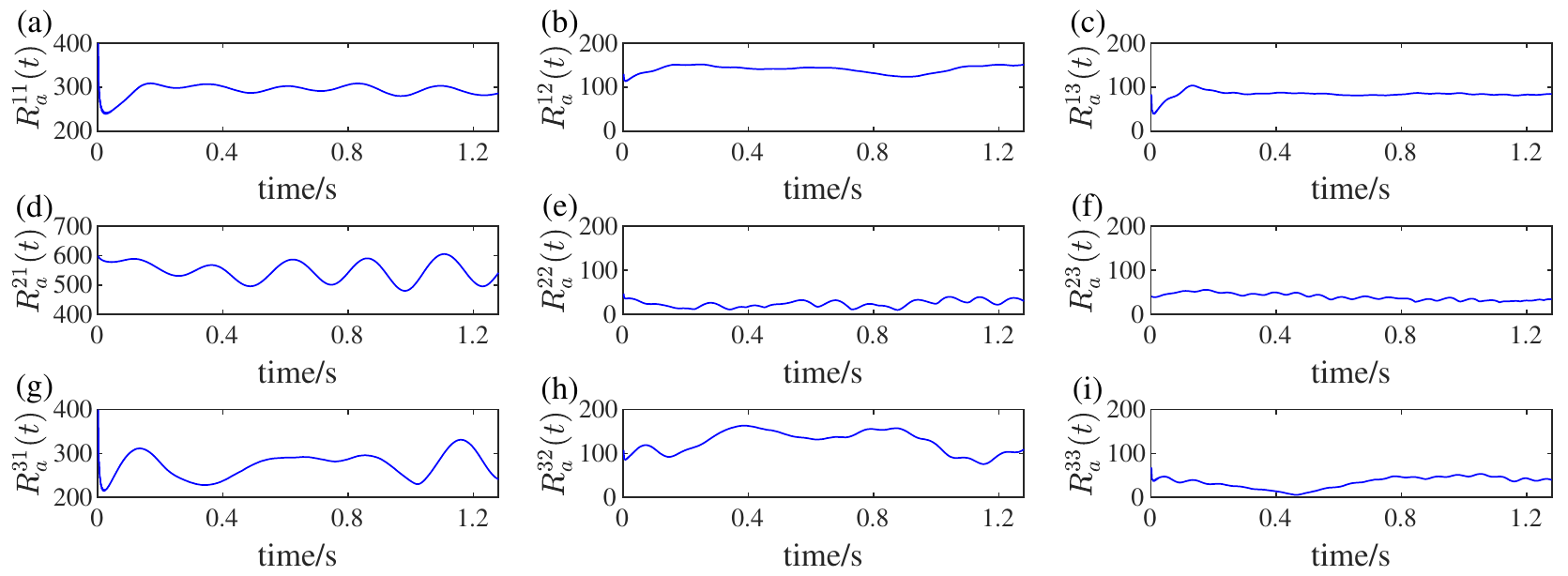}
    \caption{Semi-major axis. (a) (b) and (c) The semi-major axis of 1X, 2X, and 3X components for UGB. (d) (e) and (f) The semi-major axis of 1X, 2X, and 3X components for LGB. (g) (h) and (i) The semi-major axis of 1X, 2X, and 3X components for WGB.}
    \label{figure28}
\end{figure}
\begin{figure}
    \centering
    \includegraphics[width=0.8\linewidth]{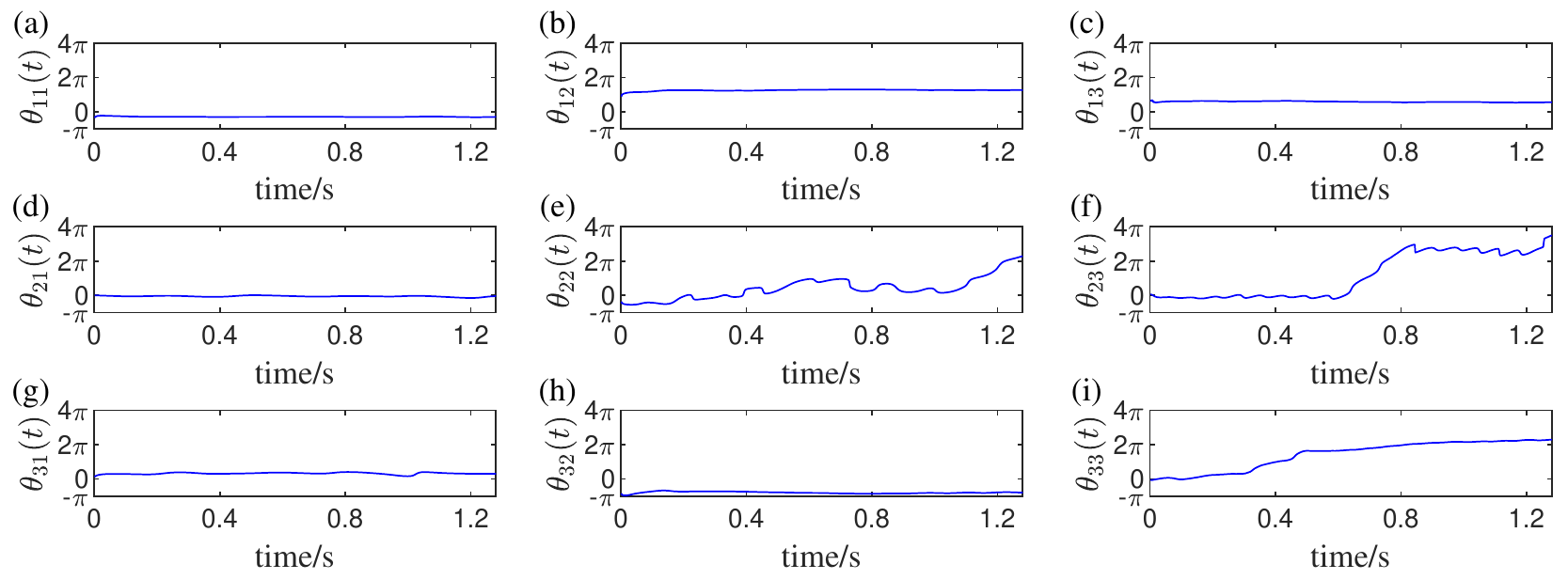}
    \caption{The instantaneous inclination angle. (a) (b) and (c) The inclination angle of 1X, 2X, and 3X components for UGB. (d) (e) and (f) The inclination angle of 1X, 2X, and 3X components for LGB. (g) (h) and (i) The inclination angle of 1X, 2X, and 3X components for WGB.}
    \label{figure29}
\end{figure}
\begin{figure}
    \centering
    \includegraphics[width=0.8\linewidth]{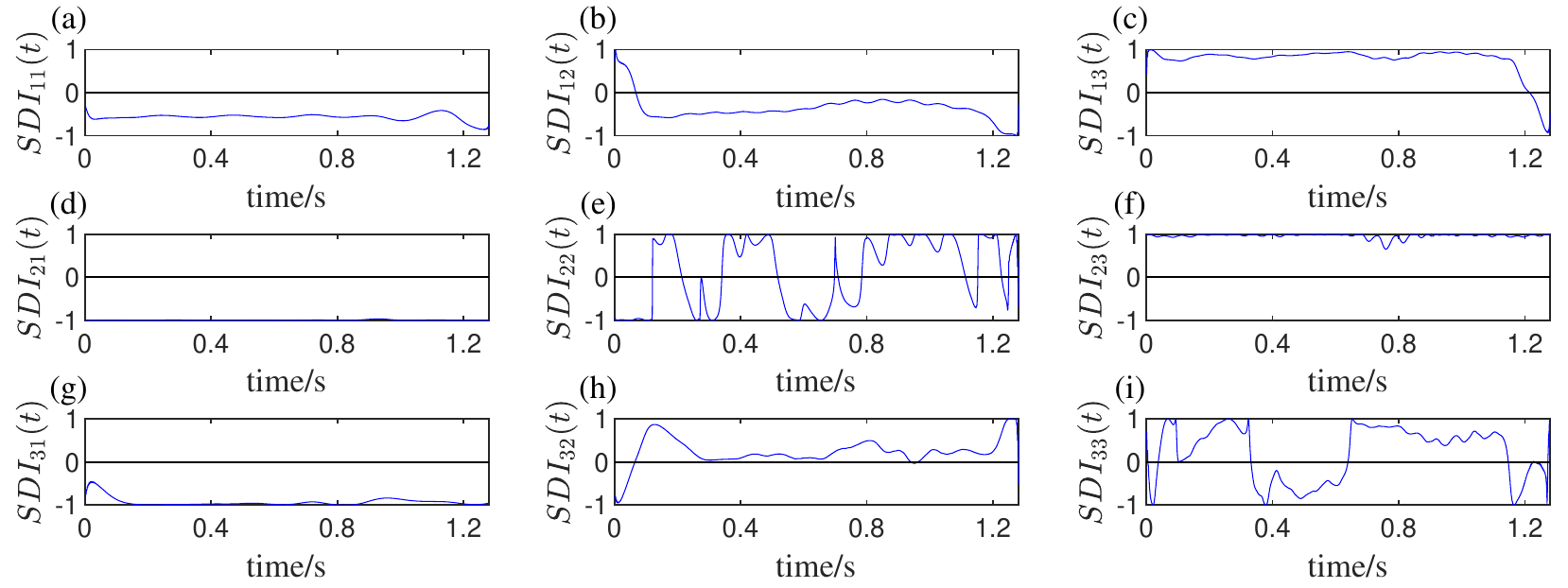}
    \caption{SDI. (a) (b) and (c) SDI of 1X, 2X, and 3X components for UGB. (d) (e) and (f) SDI of 1X, 2X, and 3X components for LGB. (g) (h) and (i) SDI of 1X, 2X, and 3X components for WGB.}
    \label{figure30}
\end{figure}
\begin{figure}
    \centering
    \includegraphics[width=1.0\linewidth]{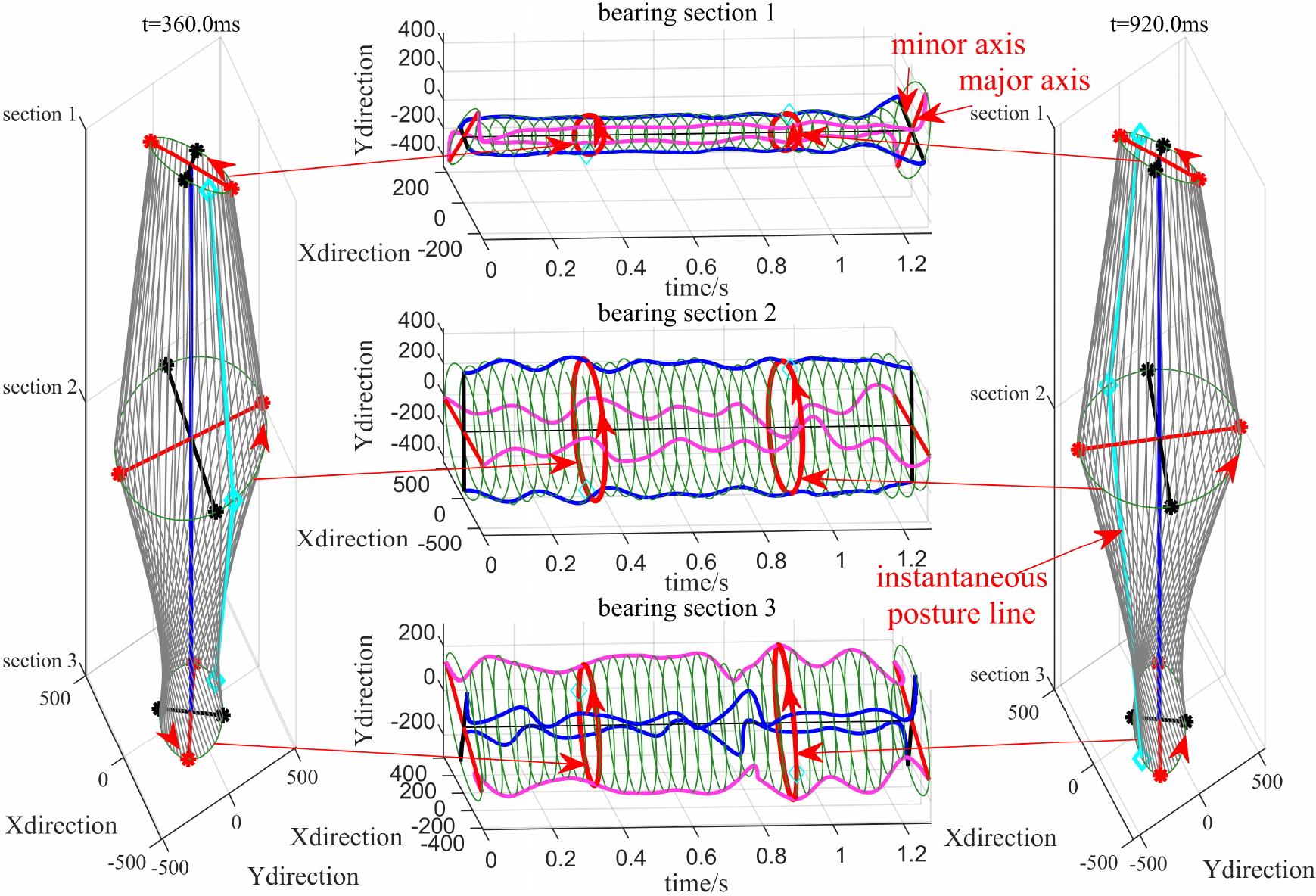}
    \caption{3D-IOM of the 1X component of the vibration signal on the three bearing sections of the rotor.}
    \label{figure31}
\end{figure}
\begin{figure}
    \centering
    \includegraphics[width=0.5\linewidth]{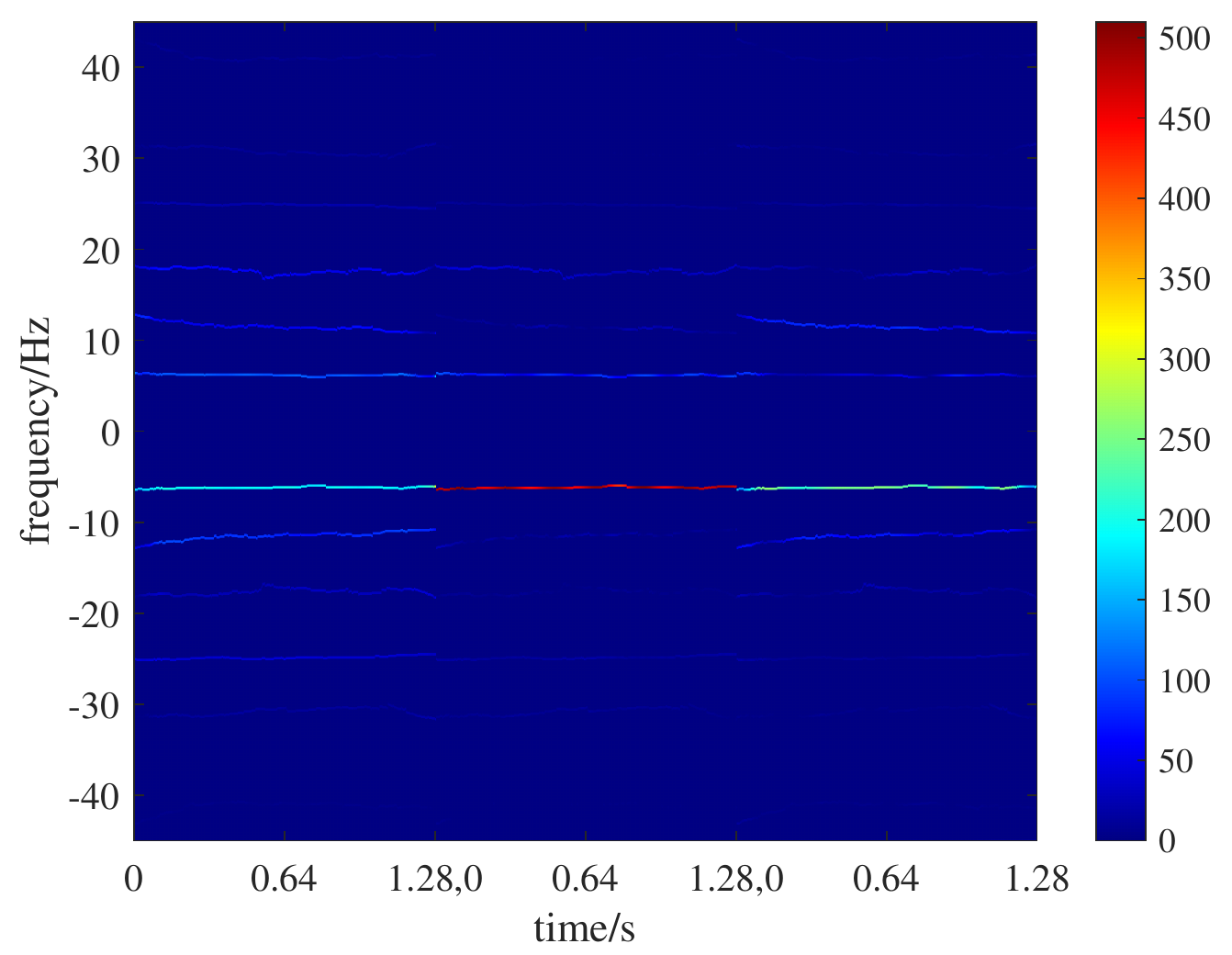}
    \caption{Time-FS of the vibration signal of turbine rotors in operation.}
    \label{figure32}
\end{figure}

The noisy signal waveform and orbit are illustrated in Fig. \ref{figure22}. The original signal had strong noise, the orbit was irregular and the noise at the WGB was the largest. The spectrum and full spectrum are shown in Fig. \ref{figure23}. The frequency spectrum indicated that each bearing section contained 7 components, so we set the modes number of MVMD to 7. The rotor orbits were all backward precession at three bearings (Fig. \ref{figure23} (c) (f) and (i)). The decomposition results processed by MVMD are presented in Fig. \ref{figure24}. From the amplitude results of the forward and backward components, it could be judged that the 1X component orbits on the three bearing sections are all backward precession, which was consistent with the results obtained from the full spectrum. Thanks to the excellent characteristics of center frequency alignment, some small amplitude component waveforms could also be extracted. For example, the 3X components at the LGB and WGB were almost a straight line. This situation was difficult to achieve by conventional methods.The reconstructed signals of 1X, 2X and 3X component are presented in Fig. \ref{figure25}, Fig. \ref{figure26}, and Fig. \ref{figure27}. It can be seen that the SNR of the reconstructed signal was significantly improved. Also, the direction of the orbit was associated with the value of SDI (see Fig. \ref{figure30}). Specifically, the value of SDI was positive when forward precession, and the angle between the major axis and the horizontal axis of the orbit should be in the range of 0 to $\frac{\pi }{2}$. Conversely, the value of SDI was negative when backward precession, and the angle between the major axis and the horizontal axis of the synthesized orbit should be in the range of $\frac{\pi }{2}$ to $\pi$. The instantaneous inclination angle is shown in Fig. \ref{figure29}. The 3D-IOM is shown in Fig. \ref{figure31}. The 3D-IOM contains more joint information and can better reflect the complex instantaneous vibration state of the rotor-bearing system. Such rich information is not available in other methods. The forward and backward components of each component of the three bearing sections are shown in Fig. \ref{figure32}.
\section{Conclusion}\label{section5}
We have introduced a novel complex-valued signal decomposition model named MCVMD, which can handle multivariate non-stationary complex-valued signals. MCVMD uses the Hilbert transform to separate the forward and backward frequency components of the signal. With the unique properties of MVMD, the signal components of multiple bearing sections of the rotor can be extracted simultaneously. Further, similar to the three-dimensional holospectrum, we used the instantaneous posture line to link the corresponding points of the instantaneous orbit at different moments. Numerical experiments illustrate the performance of the proposed method. A highlight of this paper is to visualize the instantaneous vibration state of the rotor-bearing system. The 3D-IOM could present more comprehensive information and could roughly predict the movement trend at the next moment.

Taken together, this study had gone some way towards enhancing the understanding of rotor vibration analysis under a non-stationary state and provided theoretical and practical tools for rotor a non-stationary vibration analysis. However, the main limitation of this study is that: i) The number of decomposition modes needs to be given beforehand; ii) The time-varying signal we analyzed only changes in amplitude and does not change in frequency. In future work, we will study signals that vary in both amplitude and frequency with time.

\section*{Acknowledgement}
This work was supported in part by the National Natural Science Foundation of China under Grant (Nos.51879111), in part by the Applied Fundamental Frontier Project of Wuhan Science and Technology Bureau under Grant (N0s.2018010401011269), and in part by the Hubei Provincial Natural Science Foundation of China under Grant (Nos.2019CFA068).

%% The Appendices part is started with the kcommand \appendix;


 \bibliographystyle{elsarticle-num} 
 \bibliography{reference}

\begin{thebibliography}{10}
\expandafter\ifx\csname url\endcsname\relax
  \def\url#1{\texttt{#1}}\fi
\expandafter\ifx\csname urlprefix\endcsname\relax\def\urlprefix{URL }\fi
\expandafter\ifx\csname href\endcsname\relax
  \def\href#1#2{#2} \def\path#1{#1}\fi

\bibitem{lei2013review}
Y.~Lei, J.~Lin, Z.~He, M.~J. Zuo, A review on empirical mode decomposition in
  fault diagnosis of rotating machinery, Mechanical systems and signal
  processing 35~(1-2) (2013) 108--126.

\bibitem{li2018entropy}
Y.~Li, X.~Wang, Z.~Liu, X.~Liang, S.~Si, The entropy algorithm and its variants
  in the fault diagnosis of rotating machinery: A review, IEEE Access 6 (2018)
  66723--66741.

\bibitem{al2011vibration}
F.~Al-Badour, M.~Sunar, L.~Cheded, Vibration analysis of rotating machinery
  using time--frequency analysis and wavelet techniques, Mechanical Systems and
  Signal Processing 25~(6) (2011) 2083--2101.

\bibitem{fan2008machine}
X.~Fan, M.~J. Zuo, Machine fault feature extraction based on intrinsic mode
  functions, Measurement Science and Technology 19~(4) (2008) 045105.

\bibitem{bin2012early}
G.~Bin, J.~Gao, X.~Li, B.~Dhillon, Early fault diagnosis of rotating machinery
  based on wavelet packets—empirical mode decomposition feature extraction
  and neural network, Mechanical Systems and Signal Processing 27 (2012)
  696--711.

\bibitem{hu2007fault}
Q.~Hu, Z.~He, Z.~Zhang, Y.~Zi, Fault diagnosis of rotating machinery based on
  improved wavelet package transform and svms ensemble, Mechanical systems and
  signal processing 21~(2) (2007) 688--705.

\bibitem{ying2021permutation}
W.~Ying, J.~Zheng, H.~Pan, Q.~Liu, Permutation entropy-based improved uniform
  phase empirical mode decomposition for mechanical fault diagnosis, Digital
  Signal Processing 117 (2021) 103167.

\bibitem{zhang2021novel}
Y.~Zhang, C.~Li, R.~Wang, J.~Qian, A novel fault diagnosis method based on
  multi-level information fusion and hierarchical adaptive convolutional neural
  networks for centrifugal blowers, Measurement 185 (2021) 109970.

\bibitem{gilles2013empirical}
J.~Gilles, Empirical wavelet transform, IEEE transactions on signal processing
  61~(16) (2013) 3999--4010.

\bibitem{huang1998empirical}
N.~E. Huang, Z.~Shen, S.~R. Long, M.~C. Wu, H.~H. Shih, Q.~Zheng, N.-C. Yen,
  C.~C. Tung, H.~H. Liu, The empirical mode decomposition and the hilbert
  spectrum for nonlinear and non-stationary time series analysis, Proceedings
  of the Royal Society of London. Series A: mathematical, physical and
  engineering sciences 454~(1971) (1998) 903--995.

\bibitem{dragomiretskiy2013variational}
K.~Dragomiretskiy, D.~Zosso, Variational mode decomposition, IEEE transactions
  on signal processing 62~(3) (2013) 531--544.

\bibitem{li2020novel}
Y.-x. Li, S.-b. Jiao, X.~Gao, A novel signal feature extraction technology
  based on empirical wavelet transform and reverse dispersion entropy, Defence
  Technology (2020).

\bibitem{ye2020adaptive}
X.~Ye, Y.~Hu, J.~Shen, C.~Chen, G.~Zhai, An adaptive optimized tvf-emd based on
  a sparsity-impact measure index for bearing incipient fault diagnosis, IEEE
  Transactions on Instrumentation and Measurement 70 (2020) 1--11.

\bibitem{shi2021multistage}
Y.~Shi, J.~Zhou, Multistage noise reduction processing for vibration signal of
  hydropower units, in: Journal of Physics: Conference Series, Vol. 2108, IOP
  Publishing, 2021, p. 012008.

\bibitem{yu2019novel}
H.~Yu, H.~Li, Y.~Li, Y.~Li, A novel improved full vector spectrum algorithm and
  its application in multi-sensor data fusion for hydraulic pumps, Measurement
  133 (2019) 145--161.

\bibitem{yu2020vibration}
H.~Yu, H.~Li, Y.~Li, Vibration signal fusion using improved empirical wavelet
  transform and variance contribution rate for weak fault detection of
  hydraulic pumps, ISA transactions 107 (2020) 385--401.

\bibitem{goldman1999application}
P.~Goldman, A.~Muszynska, Application of full spectrum to rotating machinery
  diagnostics, Orbit 20~(1) (1999) 17--21.

\bibitem{han1999directional}
Y.-S. Han, C.-W. Lee, Directional wigner distribution for order analysis in
  rotating/reciprocating machines, Mechanical Systems and Signal Processing
  13~(5) (1999) 723--737.

\bibitem{shravankumar2016detection}
C.~Shravankumar, R.~Tiwari, Detection of a fatigue crack in a rotor system
  using full-spectrum based estimation, Sadhana 41~(2) (2016) 239--251.

\bibitem{zhao2012multivariate}
X.~Zhao, T.~H. Patel, M.~J. Zuo, Multivariate emd and full spectrum based
  condition monitoring for rotating machinery, Mechanical Systems and Signal
  Processing 27 (2012) 712--728.

\bibitem{patel2011application}
T.~H. Patel, A.~K. Darpe, Application of full spectrum analysis for rotor fault
  diagnosis, in: IUTAM symposium on emerging trends in rotor dynamics,
  Springer, 2011, pp. 535--545.

\bibitem{jia2018coupling}
R.~Jia, F.~Ma, H.~Wu, X.~Luo, X.~Ma, Coupling fault feature extraction method
  based on bivariate empirical mode decomposition and full spectrum for
  rotating machinery, Mathematical Problems in Engineering 2018 (2018).

\bibitem{qu1989holospectrum}
L.~Qu, X.~Liu, G.~Peyronne, Y.~Chen, The holospectrum: a new method for rotor
  surveillance and diagnosis, Mechanical systems and signal processing 3~(3)
  (1989) 255--267.

\bibitem{liangsheng1998rotor}
Q.~Liangsheng, Q.~Hai, X.~Guanghua, et~al., Rotor balancing based on
  holospectrum analysis principle and practice, China mechanical engineering
  9~(1) (1998) 60--63.

\bibitem{liu2008new}
S.~Liu, L.~Qu, A new field balancing method of rotor systems based on
  holospectrum and genetic algorithm, Applied Soft Computing 8~(1) (2008)
  446--455.

\bibitem{han2011application}
J.~HAN, G.-q. ZHAO, H.~CHEN, X.-y. GONG, Application on full vector spectrum
  technology in shaft and bearing vibration signal processing, Machinery Design
  \& Manufacture 4 (2011).

\bibitem{chen2017prediction}
L.~Chen, J.~Han, W.~Lei, Z.~Guan, Y.~Gao, Prediction model of vibration feature
  for equipment maintenance based on full vector spectrum, Shock and Vibration
  2017 (2017).

\bibitem{gong2012bearing}
X.~Y. Gong, J.~Han, H.~Chen, W.~P. Lei, A bearing fault diagnosis using wavelet
  envelope spectrum based on full vector spectrum technology, in: Applied
  Mechanics and Materials, Vol. 190, Trans Tech Publ, 2012, pp. 873--879.

\bibitem{rilling2007bivariate}
G.~Rilling, P.~Flandrin, P.~Gon{\c{c}}alves, J.~M. Lilly, Bivariate empirical
  mode decomposition, IEEE signal processing letters 14~(12) (2007) 936--939.

\bibitem{rehman2009bivariate}
N.~u. Rehman, D.~Looney, T.~Rutkowski, D.~Mandic, Bivariate emd-based image
  fusion, in: 2009 IEEE/SP 15th Workshop on Statistical Signal Processing,
  IEEE, 2009, pp. 57--60.

\bibitem{yang2011bivariate}
W.~Yang, R.~Court, P.~J. Tavner, C.~J. Crabtree, Bivariate empirical mode
  decomposition and its contribution to wind turbine condition monitoring,
  Journal of Sound and Vibration 330~(15) (2011) 3766--3782.

\bibitem{tanaka2007complex}
T.~Tanaka, D.~P. Mandic, Complex empirical mode decomposition, IEEE Signal
  Processing Letters 14~(2) (2007) 101--104.

\bibitem{wang2017complex}
Y.~Wang, F.~Liu, Z.~Jiang, S.~He, Q.~Mo, Complex variational mode decomposition
  for signal processing applications, Mechanical systems and signal processing
  86 (2017) 75--85.

\bibitem{ur2009empirical}
N.~ur~Rehman, D.~P. Mandic, Empirical mode decomposition for trivariate
  signals, IEEE Transactions on signal processing 58~(3) (2009) 1059--1068.

\bibitem{rehman2010multivariate}
N.~Rehman, D.~P. Mandic, Multivariate empirical mode decomposition, Proceedings
  of the Royal Society A: Mathematical, Physical and Engineering Sciences
  466~(2117) (2010) 1291--1302.

\bibitem{ahrabian2015synchrosqueezing}
A.~Ahrabian, D.~Looney, L.~Stankovi{\'c}, D.~P. Mandic, Synchrosqueezing-based
  time-frequency analysis of multivariate data, Signal Processing 106 (2015)
  331--341.

\bibitem{ur2019multivariate}
N.~ur~Rehman, H.~Aftab, Multivariate variational mode decomposition, IEEE
  Transactions on Signal Processing 67~(23) (2019) 6039--6052.

\bibitem{cui2020instantaneous}
X.~Cui, C.~Li, B.~Li, Y.~Li, Instantaneous feature extraction and
  time-frequency representation of rotor purified orbit based on vold-kalman
  filter, IEEE Transactions on Instrumentation and Measurement (2020).

\bibitem{bachschmid2004diagnostic}
N.~Bachschmid, P.~Pennacchi, A.~Vania, Diagnostic significance of orbit shape
  analysis and its application to improve machine fault detection, Journal of
  the Brazilian Society of Mechanical Sciences and Engineering 26~(2) (2004)
  200--208.

\bibitem{lee1998directional}
C.-W. Lee, Y.-S. Han, The directional wigner distribution and its applications,
  Journal of Sound and Vibration 216~(4) (1998) 585--600.

\bibitem{qu2007holospectrum}
L.~Qu, Holospectrum and holobalancing technique in machinery diagnosis, Science
  Publication House (2007).

\end{thebibliography}


\end{document}